\documentclass{emulateapj}

\newcommand{\spitzer}{{\it Spitzer}}

\newcommand{\um}{$\mu$m}
\newcommand{\hii}{{\ion{H}{2}~}}

\newcommand{\degree}{^{\circ}}
\newcommand{\Msun}{M$_{\sun}$}
\newcommand{\Lsun}{L$_{\sun}$}

\slugcomment{For Publication in the Chandra Carina Complex Project Special Issue of the Astrophysical Journal Supplement Series}

\shorttitle{Candidate X-ray-Emitting OB Stars in Carina}
\shortauthors{Povich et al.}

\begin{document}

\title{Candidate X-ray-Emitting OB Stars in the Carina Nebula Identified Via Infrared Spectral Energy Distributions}

\author{Matthew S. Povich,\altaffilmark{1,2}
  Leisa K. Townsley,\altaffilmark{1} 
  Patrick S. Broos,\altaffilmark{1} Marc Gagn\'{e},\altaffilmark{3} 
  Brian L. Babler,\altaffilmark{4} R\'{e}my Indebetouw,\altaffilmark{5} 
  Steven R. Majewski,\altaffilmark{5} Marilyn R. Meade,\altaffilmark{4}
  Konstantin V. Getman,\altaffilmark{1}
  Thomas P. Robitaille,\altaffilmark{6,7}
  \& Richard H. D. Townsend\altaffilmark{4}
}

\altaffiltext{1}{Department of Astronomy and Astrophysics, The Pennsylvania
  State University, 525 Davey Laboratory, University Park, PA 16802}
\altaffiltext{2}{NSF Astronomy and Astrophysics Postdoctoral Fellow; povich@astro.psu.edu}
\altaffiltext{3}{Department of Geology and Astronomy, West Chester University, West Chester, PA 19383}
\altaffiltext{4}{Department of Astronomy, University of Wisconsin, 475
N. Charter Street, Madison, WI 53706}
\altaffiltext{5}{Department of Astronomy, University of Virginia, P. O. Box 400325, Charlottesville, VA 22904-4325}
\altaffiltext{6}{Harvard--Smithsonian Center for Astrophysics, 60 Garden Street, Cambridge, MA 02138}
\altaffiltext{7}{{\it Spitzer} Postdoctoral Fellow}

\begin{abstract}

We report the results of a new survey of massive, OB stars throughout
the Carina Nebula using the X-ray point source catalog provided by the
{\it Chandra} Carina Complex Project (CCCP) in conjunction with
infrared (IR) photometry from the Two Micron All-Sky Survey and the
{\it Spitzer Space Telescope} Vela--Carina survey.  Mid-IR photometry
is relatively unaffected by extinction, hence it provides strong
constraints on the luminosities of OB stars, assuming that their
association with the Carina Nebula, and hence their distance, is
confirmed.  We fit model stellar atmospheres to the optical ($UBV$)
and IR spectral energy distributions (SEDs) of 182 OB stars with known
spectral types and measure the bolometric luminosity and extinction
for each star.  We find that the extinction law measured toward the OB
stars has two components: $A_V=1$--1.5~mag produced by foreground dust
with a ratio of total-to-selective absorption $R_V=3.1$ plus a
contribution from local dust with $R_V>4.0$ in the Carina molecular
clouds that increases as $A_V$ increases.  Using X-ray emission as a
strong indicator of association with Carina, we identify 94 candidate
OB stars with $L_{\rm bol}\ga 10^4$~\Lsun\ by fitting their IR SEDs.
If the candidate OB stars are eventually confirmed by follow-up
spectroscopic observations, the number of cataloged OB stars in the
Carina Nebula will increase by ${\sim}50\%$. Correcting for
incompleteness due to OB stars falling below the $L_{\rm bol}$ cutoff
or the CCCP detection limit, these results potentially double the size
of the young massive stellar population.
\end{abstract}


\section{Introduction}

The Great Nebula in Carina (NGC 3372) hosts
200 known OB stars \citep[][hereafter G11]{NS06a,OBCat}, 
the largest collection of young massive stars within 3~kpc of the Sun. This population
includes some of the most massive stars ever discovered, including the
famous Luminous Blue Variable $\eta$ Carinae \citep{DH97}. 
The 2.3~kpc distance to the Carina Nebula has been measured accurately from the
expansion of the Homunculus Nebula produced by $\eta$~Car \citep{AH93,NS06b},
and the principal ionizing star
clusters, Trumpler (Tr) 16 and Tr 14, suffer relatively low
dust extinction \citep[$A_V\la 2$~mag;][]{DGE01,JA07,CCCPhawki}. 
Carina therefore offers unique observational advantages for
multiwavelength studies of a well-resolved population of massive
stars, providing a large sample of young OB stars 
in a single giant molecular cloud complex.

There is reason to suspect that significant numbers of
massive stars have yet to be cataloged in Carina. 
The Carina Nebula extends ${>}1\degree$ across the sky, hence
spatially unbiased observations
of the entire complex are challenging. While there are relatively few
bright infrared (IR) sources consistent with embedded massive stars 
\citep{JR04,SB07}, OB stars that have shed their natal
envelopes could be hiding in 
regions of higher extinction: behind molecular cloud fragments
\citep{YY05}, 
in the dark, V-shaped dust lane south of Tr 16 seen in visible-light
images \citep{SB07}, 
or intermingled with embedded YSOs in regions 
of active star formation such as the South Pillars \citep[][hereafter
P11]{TM96,JR04,spitzcar,Paper I}. 
New surveys from the {\it Chandra X-ray Observatory} and the {\it
  Spitzer Space Telescope} offer wide-field views encompassing
the bulk of the young stellar population in Carina. Compared to the
optical studies historically used to catalog OB stars in Carina, X-ray and  
IR observations are much less affected by extinction.  
These surveys provide an opportunity to
carry out a comprehensive search for OB stars.

In this paper, we present the results of a survey of candidate OB
stars throughout the Carina Nebula. In the process, we further develop
the spectral energy distribution (SED) fitting method pioneered by 
\citet{WP08} 
to identify candidate massive stars and constrain their properties. In \S2
we briefly summarize the observations and basic data analysis. We
describe our SED fitting methodology and apply it to a
``validation sample'' of Carina OB stars with known spectral types in
\S3. In \S4 we extend the SED fitting method to identify a sample of
candidate X-ray-emitting OB stars. We discuss and
summarize our results in \S5.

\section{Observations and Basic Data Analysis}

We used data products from the {\it Chandra} Carina Complex
Project \citep[CCCP;][]{CCCPintro}, Two-Micron All Sky Survey
\citep[2MASS;][]{2MASS}, and the 
\spitzer\ Vela--Carina Survey (PI S. R. Majewski), as described in
P11. P11 constructed a catalog of
young stellar objects (YSOs) identified on
the basis of IR excess
emission in the Vela--Carina Point Source Catalog; as a first step,
SED fitting was used to 
filter out 50,586 of 54,155 
sources in the CCCP survey area, those that were consistent
with stellar photospheres reddened by interstellar 
dust alone (no IR excess emission from circumstellar dust).
This large sample of stars without IR excess, discarded from the
analysis of P11, 
forms the basis of our current study.
We use the $\chi^2$-minimization SED fitting tool of \citet{fitter},
which applies interstellar 
reddening directly to the stellar atmosphere models \citep{Kurucz}
before fitting 
them to the data, hence incorporating the optical extinction $A_V$
as a free parameter. The fitting tool can accept any extinction law,
defined as opacity as a function of wavelength. We employed a
\citet{CCM89} extinction law, characterized by the ratio of total to
selective absorption $R_V$, for the optical--ultraviolet (UV) part of
the spectrum 
and the \citet{I05} extinction law in the IR.

P11 also identified stars with
``marginal'' IR excesses in the 5.8 or 8.0~\um\ bands of the \spitzer\
Infrared Array Camera \citep[IRAC;][]{IRAC} and discarded them as unreliable
YSO candidates. The SEDs of these sources are well-fit by stellar
atmosphere models (see P11 for the definition of ``well-fit'') when
the IR photometric band(s) affected by potential excess are excluded
from the SED fitting procedure. We incorporated these marginal excess
stars into the 
present study, suppressing fitting to the band(s) affected by apparent
excess emission. 

Compared to traditional photometric analysis techniques using
combinations of color-color and color-magnitude diagrams, the SED
fitting approach offers the considerable advantage
of analyzing all available
photometric data simultaneously \citep{CHORIZOS}.
By analyzing the sets of models that fit up to $N_{\rm data}=10$
photometric datapoints
spanning the optical ($UBV$), near-IR 
(2MASS $JHK_s$), and mid-IR (IRAC) spectrum, we can
probe interrelationships of various physical 
parameters, such as extinction and luminosity, and test which
wavelengths are most important for 
constraining a given parameter. 
We define the set $i$ of well-fit models for each source as
$\chi_i^2-\chi_0^2\le 2N_{\rm data}$, where $\chi^2_i$ and $\chi^2_0$
are the goodness-of fit parameters for the $i$th and best-fit models,
respectively and $N_{\rm data}$ is the number of photometric
datapoints available for each source \citep[][P11]{P09}. This
criterion is deliberately liberal, typically returning 
${\sim}10^3$ well-fit stellar atmosphere models to each source to
explore the full range of potential parameter space.
Another benefit of the SED fitting approach is a decreased dependence of
the fitting results on 
any single photometry band; the impact of measurement
uncertainties affecting specific bands is reduced by the inclusion of more data. 

\section{Validation Sample: Spectroscopically Confirmed OB Stars}

G11 
compiled a catalog of 200 OB stars in
Carina with spectral types known from spectroscopy \citep[e.g.,][and
references therein]{Skiff09} 
and available $UBV$
photometry.
In this catalog, 182 stars have 2MASS+IRAC photometry
in the Vela--Carina Point-Source Archive of sufficient quality
for our SED analysis, meaning the sources showed no significant
variability\footnote{The epochs of the 2MASS and Vela--Carina
observations were separated by years.} and were not saturated, confused,
or swamped by nebular emission. These 182 stars form the validation sample
(Table~1) for our SED fitting method. We were able to fit the
full SED from $U$ to 
8.0~\um\ in the majority of cases. Because the $UBV$ photometry was
obtained from various sources, many of which did not publish
photometric uncertainties, we conservatively set the
uncertainties on the optical flux densities to 20\% (${\sim}0.2$~mag)
for the SED fitting (the 2MASS and IRAC photometry constitute a
homogeneous dataset, and we
used an uncertainty floor of 10\% for the IR datapoints, see P11). 
The SEDs of stars 
undetected at 8 \um\ 
or exhibiting marginal IR excesses can generally
be fit well with stellar atmosphere models for $\lambda \le 4.5$~\um.

Among the validation sample, 22 stars show excess emission in one or more
IRAC bands. This is not necessarily evidence for circumstellar disks;
the apparent excess 
could instead be due to (1) systematic photometric errors \citep{P09}, (2) dust
trapped near the star in a bow shock or remnant envelope\footnote{Some
  of the ``extended red objects'' identified by \citet{spitzcar} fall
  into this category.} 
\citep{P08}, 
or (3) confusion with a lower-mass star with its own IR excess. In all
cases except (1) this excess emission is astrophysically interesting
and may be correlated with the youngest OB stars, so 
we note sources with potential IR excess in column (14) of Table~1.

\subsection{Luminosity and Extinction Derived from SED Fitting\label{lbolavsed}}

In general, hundreds of different atmosphere models can be fit to a
given SED, spanning a wide range in stellar effective temperature
$T_{\rm eff}$. This is due to a degeneracy between $T_{\rm
  eff}$ and $A_V$ that can be resolved with additional information.
Using the known spectral types (ST) for the 182 stars in the
validation sample (G11), 
we assigned $T_{\rm eff}$(ST) using
the calibrations of $T_{\rm eff}$ versus spectral type from 
\citet{MSH05} 
for O stars and the extrapolation provided by
\citet{PC05} for early B stars. 
With the parameter $T_{\rm eff}=T_{\rm eff}$(ST) fixed,
the bolometric
luminosity $L_{\rm bol}$ and extinction $A_V$ 
become uniquely determined. 

%
\begin{figure*}
\plottwo{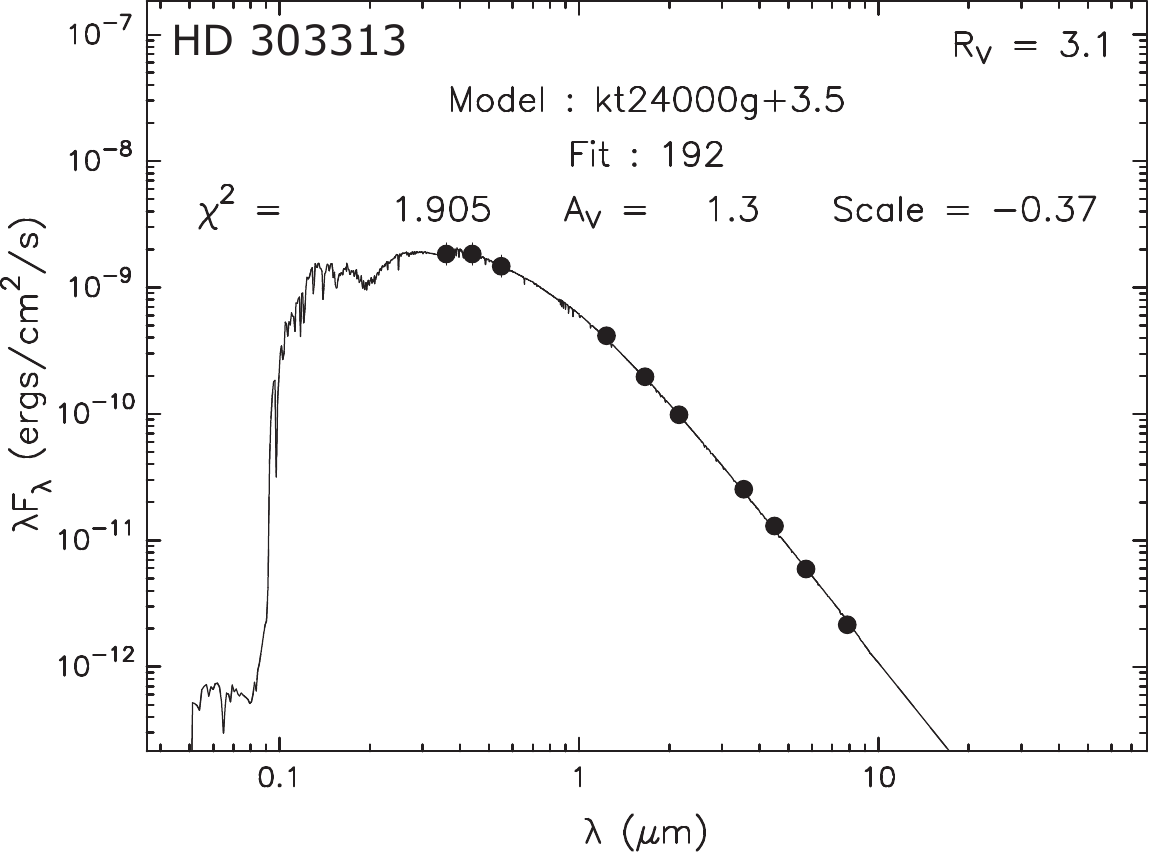}{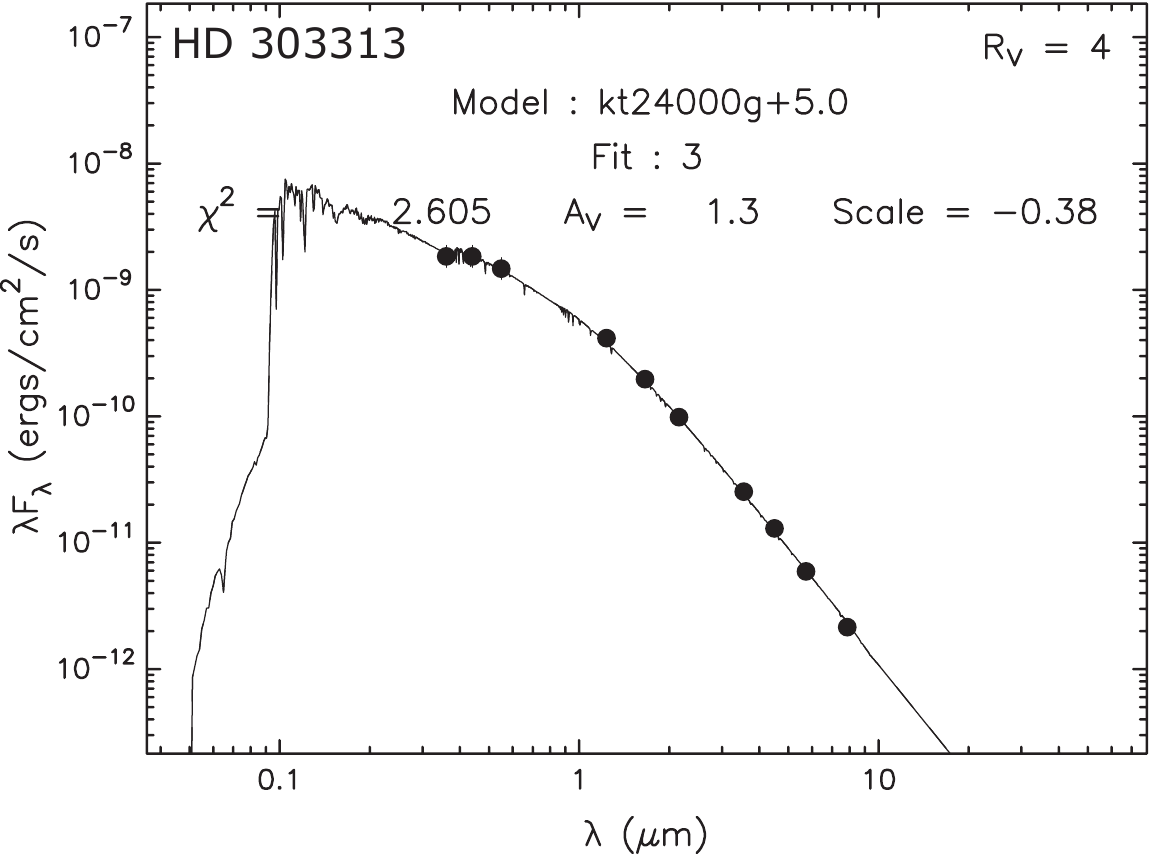}\\
\plottwo{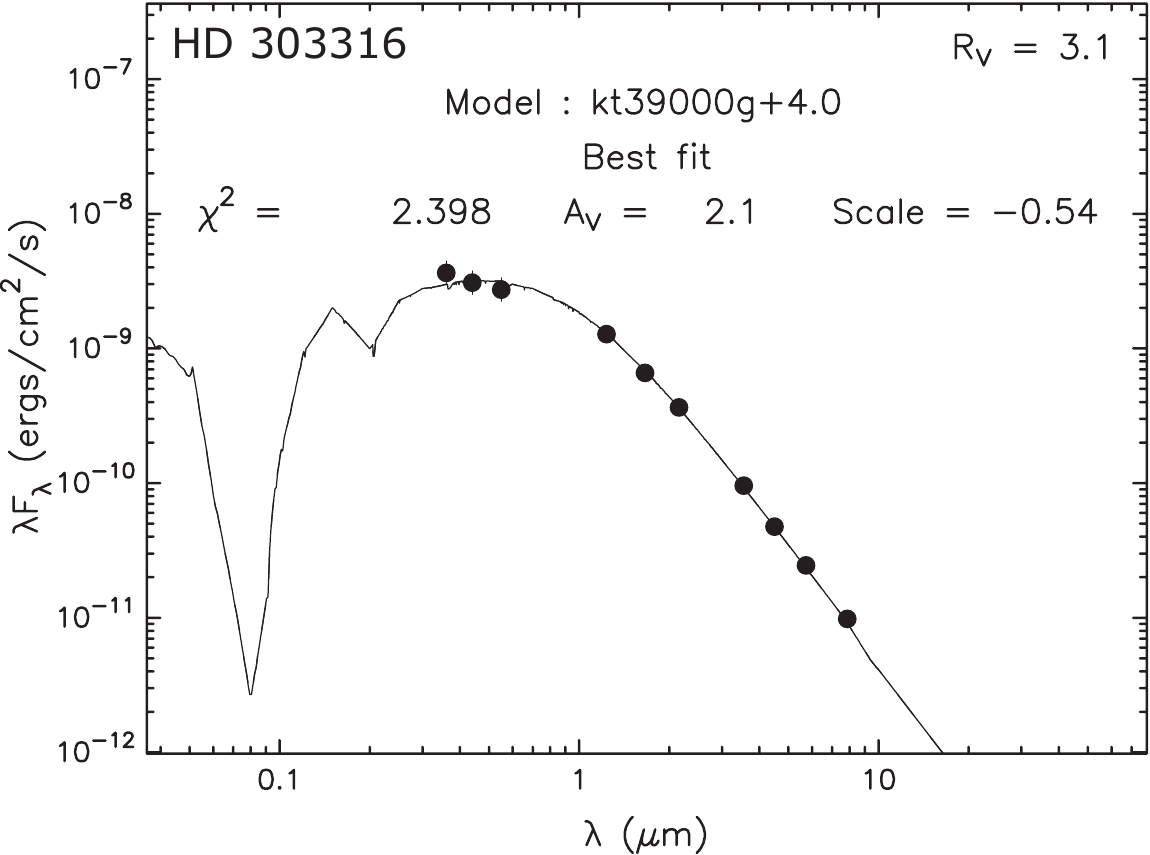}{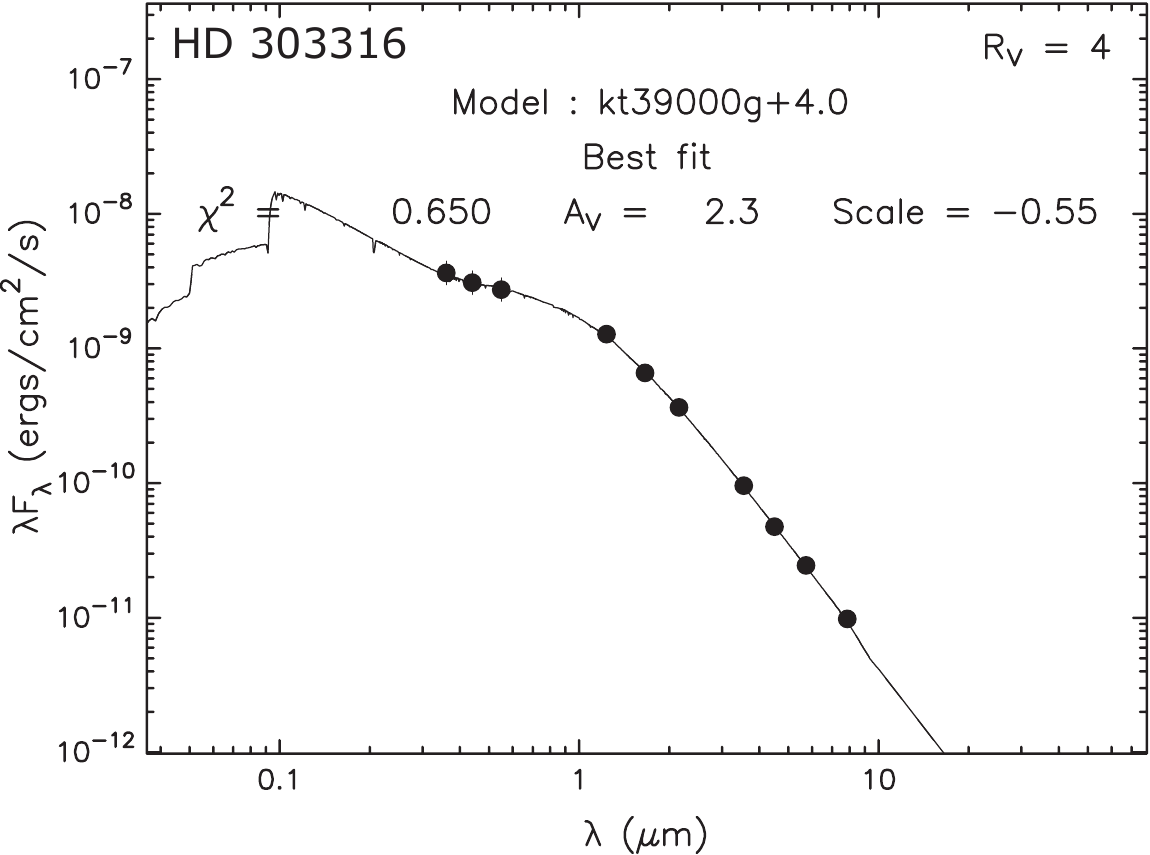}\\
\plottwo{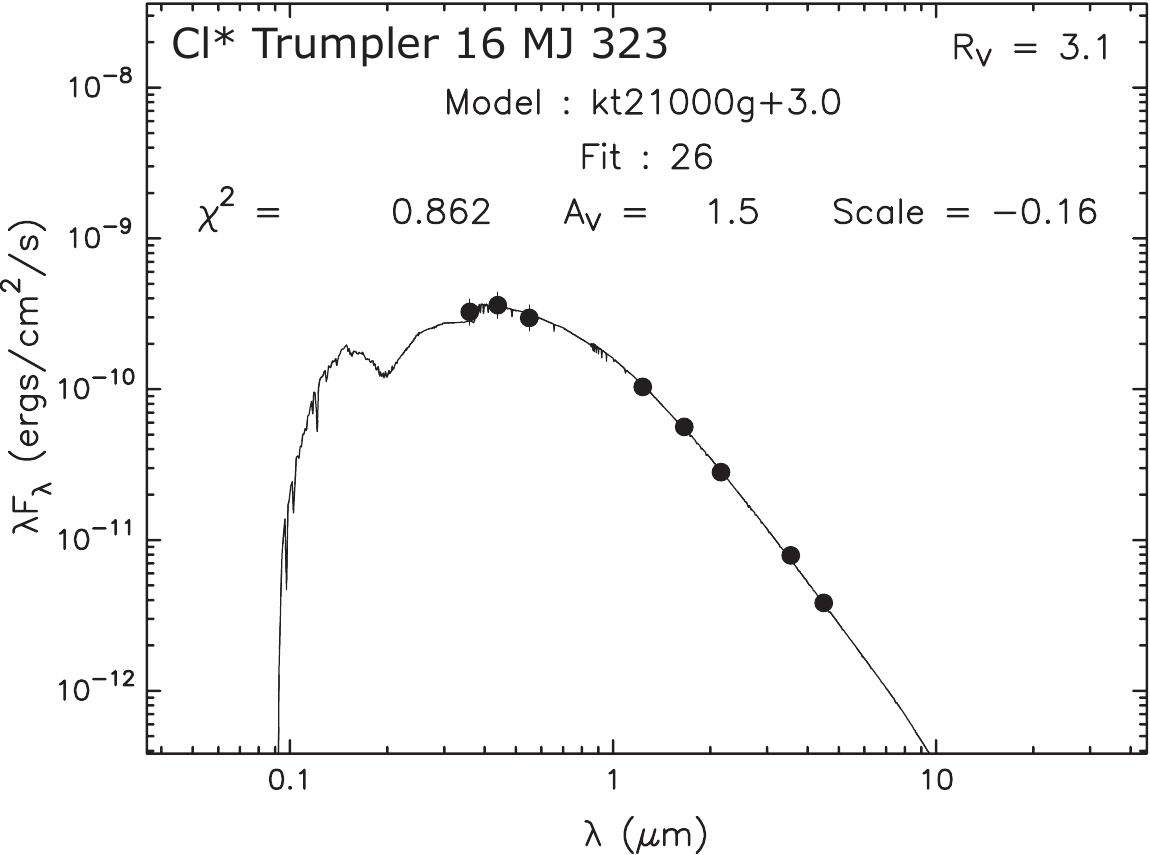}{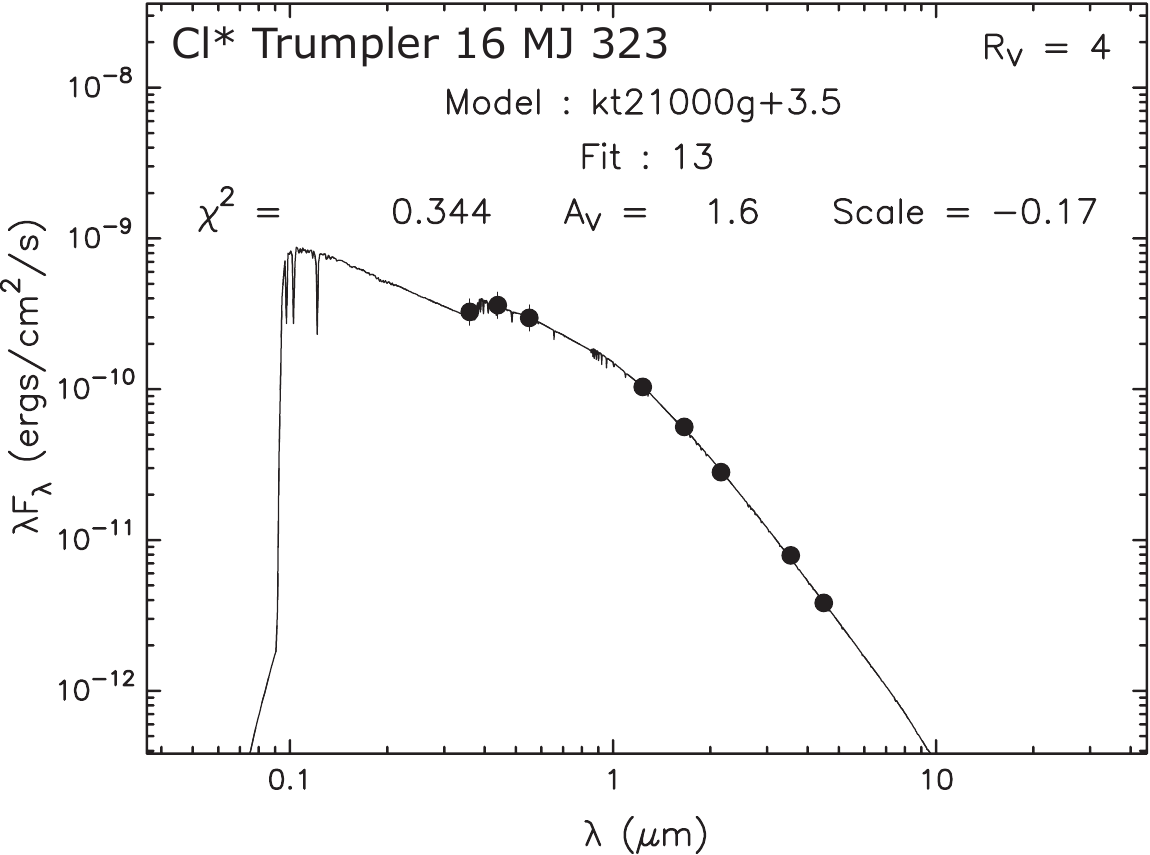}\\
\caption{Example plots illustrating the fitting of reddened ATLAS9 model
  atmospheres (curves) to the optical through IR SEDs (dots) of 3 OB
  stars in Carina with known spectral types (G11). A single well-fit model
  with effective temperature $T_{\rm eff}$ corresponding to the spectral
  type is displayed in each panel, along with the model and fit
  parameters. Each star was fit using two different extinction laws
  (see \S\ref{rv});
  $R_V=3.1$ (left column) and $R_V=4.0$ (right column).
\label{SEDs}}
\end{figure*}
%
We then chose SED fits corresponding to that
temperature for subsequent analysis. Example plots of ATLAS9 
stellar atmospheres \citep{Kurucz} fit to the SEDs of 3 validation
stars are given in Figure~\ref{SEDs}, for 2 different extinction
laws (see \S\ref{rv} below). In each plot, a single example from the
family of 
well-fit  
model atmospheres is shown, chosen to show a model 
that matches $T_{\rm eff}$(ST) for
each star (G11): HD 303313 (B1.5 V, 24,000~K), HD 303316 (O6 V, 39,000~K), and 
Cl* Trumpler 16 MJ 323 (Tr16-18; B2 V, 21,000~K). 
In addition to temperature, each model atmosphere is specified by
surface gravity and metallicity. The metallicity parameter was
ignored in our analysis, since it primarily affects
spectral lines and has negligible impact on the SED shapes. The impact
of surface gravity and the choice of stellar atmosphere models on the
SED shapes is discussed in \S\ref{sys} below.

Information about the SED fitting results is shown in each panel of
Figure~\ref{SEDs}.   
The goodness-of-fit parameter $\chi^2$ is used simply as a binary discriminator
separating well-fit from poorly-fit models (see P11 for details). 
Assuming $d=2.3$ kpc, the geometrical scale factor, defined
as $\log{d/R_{\star}}$, 
gives the stellar radius $R_{\star}$, which can be converted straightforwardly
to $L_{\rm bol}$ as a function of $T_{\rm eff}$.
With $T_{\rm eff}$ fixed, the $A_V$ parameter strongly influences the SED shape.
The effect of the assumed extinction law is thus apparent in
Figure~\ref{SEDs}: $R_V=3.1$ produces better results for the
least-reddened example star, HD 303313 ($A_V=1.3$~mag), in terms of
both the visual 
agreement between the model and the data and the $\chi^2$ of the fits;
while conversely $R_V=4.0$ works better for the most-reddened star, HD
303316 ($A_v=2.3$~mag). Cl* Trumpler 16 MJ 323 
represents an intermediate case with $A_V=1.6$, where the visual
quality of the fits is similar between the two extinction
laws, but $\chi^2$ marginally favors $R_V=4.0$.

%
\begin{figure}
\epsscale{1}
\plotone{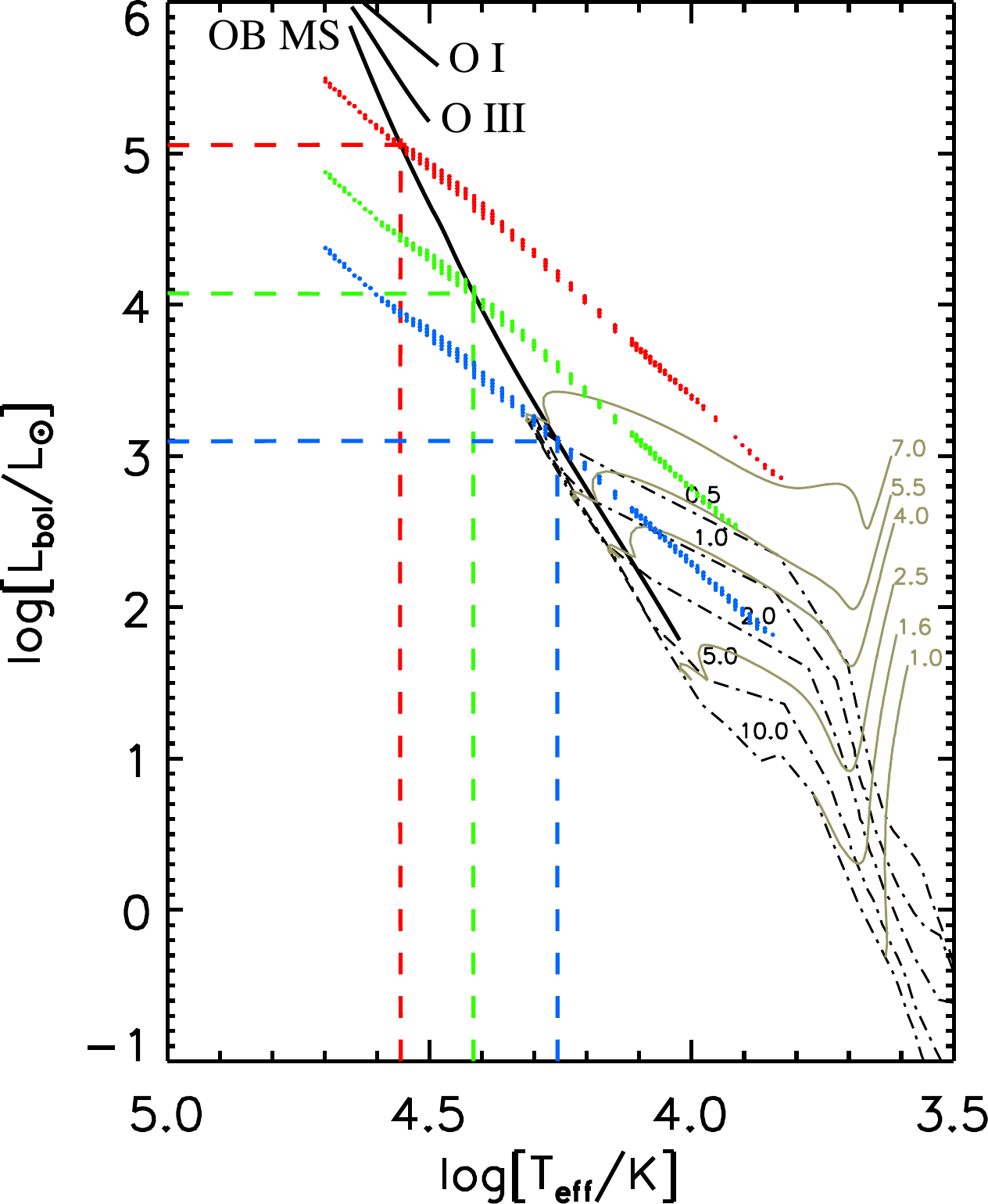}\\
\plotone{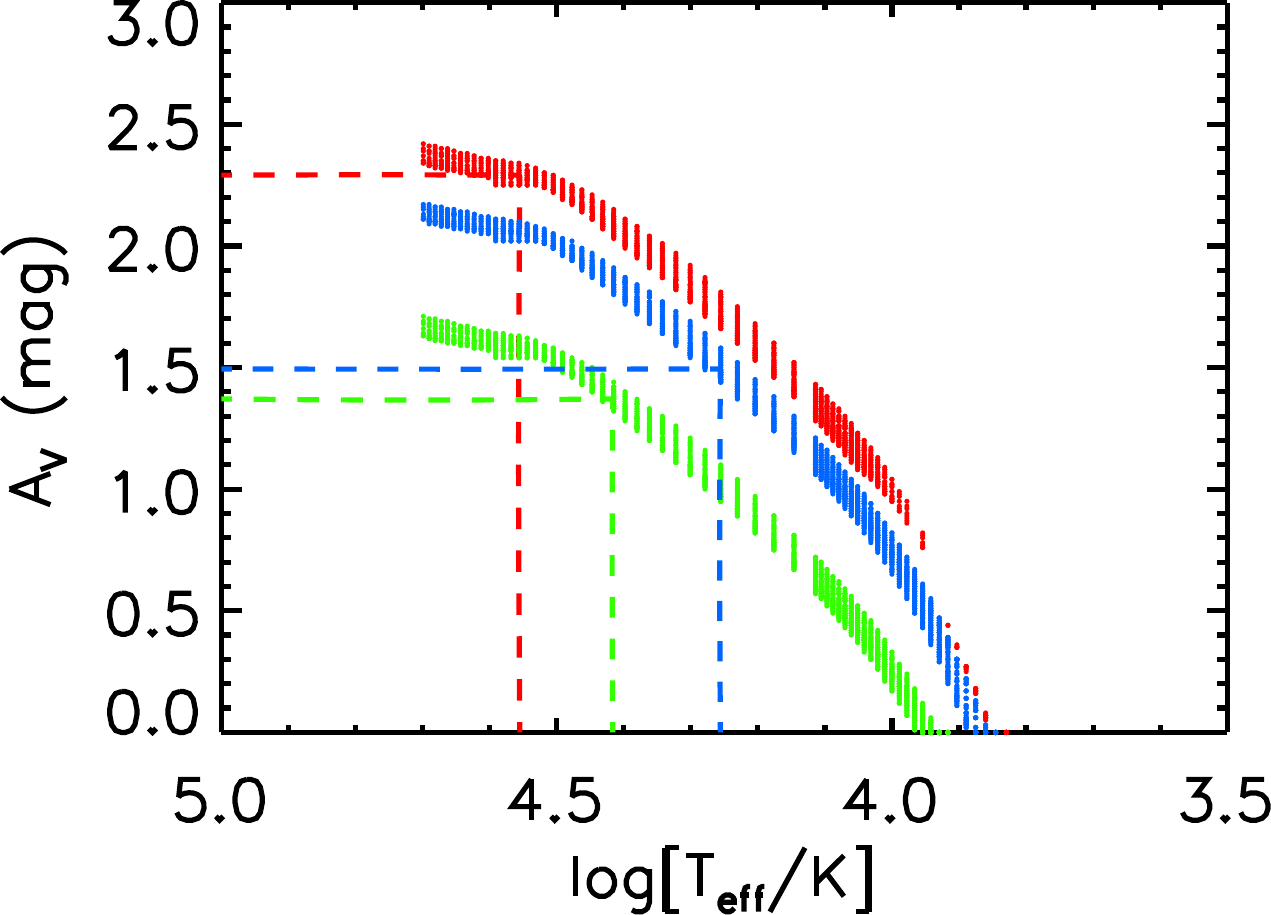}
\caption{{\it Top:} Theoretical H-R diagram with the loci of {\it all}
  well-fit SED models plotted for the 3 stars shown in
  Figure~\ref{SEDs} (colored dots: HD 303313 = green, HD 303316 = red,
  Cl* Trumpler 16 MJ 323 = blue). The quantization 
  in $T_{\rm eff}$ reflects the gridding of the ATLAS9
  atmospheres \citep{Kurucz}. Solid black curves show the locations of the
  theoretical OB MS, O giants, and O 
  supergiants \citep{MSH05,dJN87}. 
  PMS tracks from \citet{SDF00} 
  in the range of 1.0 to 7.0~\Msun\ are plotted
  as gray curves, and isochrones from 0.5 to 10.0 Myr are plotted as
  black dash-dotted curves. The intersection of each locus with the MS
  defines the maximum physically plausible temperature $T_{\rm
    eff}^{\rm MS}$ and
  bolometric luminosity $L_{\rm bol}^{\rm MS}$ (colored, dashed lines).
{\it Bottom:} Loci of $A_V$ versus $T_{\rm eff}$ for the same 3
stars (assuming $R_V=4.0$). The maximum interstellar extinction
$A_V^{\rm MS}$
for each star corresponds to 
 $T_{\rm eff}^{\rm MS}$.
\label{HRD}}
\end{figure}
%
In a traditional photometric study, all stars are plotted
simultaneously on a color-magnitude diagram; 
age, luminosity, and reddening are inferred for the ensemble by comparing
loci of stars with the positions of reddening vectors, theoretical
isochrones, and/or 
evolutionary tracks. The SED fitting method enables us to turn this
approach around, instead projecting the 3-dimensional locus of the
family of well-fit model 
parameters ($T_{\rm eff},L_{\rm bol},A_V$) for {\it each} star
onto a {\it theoretical} Hertzsprung-Russell (H-R) diagram
(Figure~\ref{HRD}, top). Models placing a star to the left of the
theoretical main
sequence (MS; as defined by Martins et al.~2005 for O stars, Crowther
2005 for early B stars, and de Jager \& Nieuwenhuijzen 1987 for late B
stars) can be discarded
as unphysical. A companion plot of $A_V$ against $T_{\rm
  eff}$ (Figure~\ref{HRD}, bottom) represents a second plane in the 3-D
parameter space and illustrates the degeneracy between
higher extinction and higher temperature. Since the {\it maximum}
allowed $T_{\rm eff}$ corresponds to the MS, the
maximum $A_V^{\rm MS}$ corresponds to $T_{\rm eff}^{\rm MS}$.
If $T_{\rm eff}$(ST) is known from spectroscopy, then 
$L_{\rm bol}$ and $A_V$ can
{\it always} be determined uniquely.
Values 
of $\log{L_{\rm bol}}$ and $A_V$(SED) for all 182 stars in the
validation sample are reported in columns (4) and (5) of Table~1; these
 parameter values were computed for each star by slicing the locus of model fits
{\it at the adopted spectral type temperature}, $T_{\rm
  eff}$(ST) in column (3), and taking the mean along both the $L_{\rm bol}$ and $A_V$ dimensions. 

CCCP, the Vela--Carina Survey, and 2MASS provide
homogeneous X-ray and IR coverage spanning the Carina Nebula, but
unfortunately no single complementary dataset exists that
gives high-quality optical photometry for the entire CCCP survey
area. $UBV$ photometry for the 
validation sample was 
compiled from heterogeneous sources in the literature by G11. 
It is not feasible to incorporate optical photometry into an
unbiased search for massive stars throughout Carina, some of which 
could be highly obscured at optical wavelengths. We therefore
fit the SEDs of the stars in the
validation sample a second time, discarding the $UBV$ photometry, to
test how well SEDs from IR photometry alone constrain the derived
physical properties. 

%
\begin{figure}
\epsscale{1.0}
\plotone{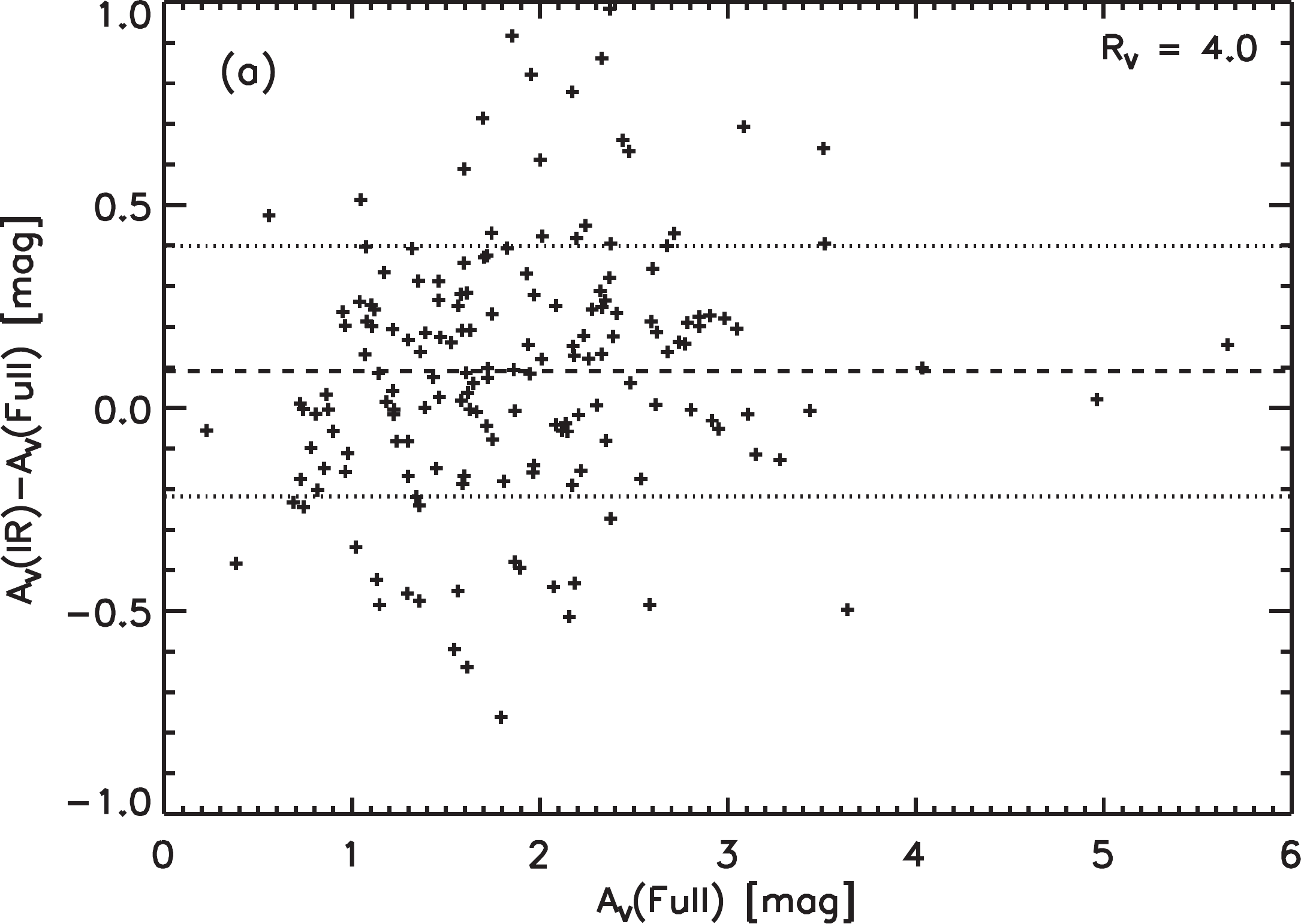}\\
\plotone{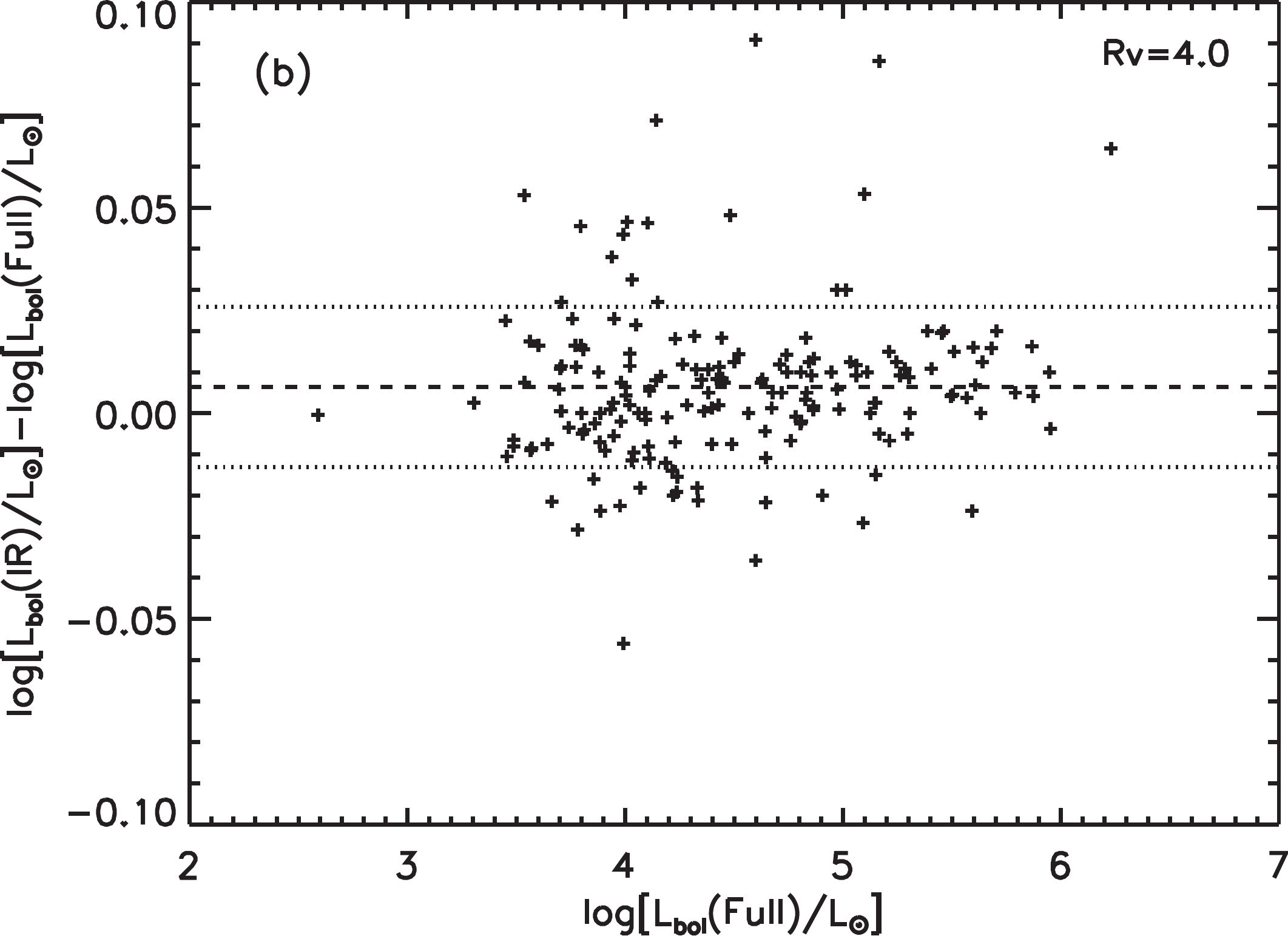}
\caption{Difference in derived interstellar extinction (a) and bolometric
  luminosity (b) for the validation sample obtained by fitting
  optical+IR (Full) and IR-only (IR) SEDs. Dashed lines show the mean and dotted lines show the
  ${\pm}1\sigma$ scatter.
\label{IRonly}}
\end{figure}
In Figure~\ref{IRonly} we plot the differences in $A_V$(SED) and $\log{L_{\rm
    bol}}$ 
 produced by comparing the
results from fitting {\it only} the IR
portions of the SEDs 
to the results from fitting the {\it full}
optical+IR SEDs 
for all 182 stars in the validation sample. 
The $1\sigma$ scatter
in $A_V{\rm (IR)}-A_V{\rm (Full)}$ is ${\sim}0.3$~mag
(Figure~\ref{IRonly}a). The mean offset of ${\sim}0.1$ mag is
not significant, given the scatter. 
The scatter in $\log{L_{\rm
    bol}{\rm (IR)}}-\log{L_{\rm bol}{\rm (Full)}}$ is ${<}0.02$~dex
(note the axis scale in Figure~\ref{IRonly}b), hence the $L_{\rm bol}$
values derived from fitting only the IR portion of the SED generally reproduce
the full SED results to within 5\%. 
Since extinction in the IR
is negligible in comparison to visible light, the IR SED sets 
the overall scaling of the spectrum, determining $L_{\rm
  bol}$. 
Optical photometry then
provides a {\it direct} measure of $A_V$ (assuming $T_{\rm eff}$ is
known, see Figure~\ref{HRD}). The majority of stars in the validation
sample have 
$A_V<3$ mag (Table~1 and Figure~\ref{IRonly}a), or $A_K<0.37$
mag \citep{CCM89}, and it is not surprising that the IR SEDs alone
do not tightly constrain $A_V$, given the conservative estimates of
photometric uncertainty used in the SED fitting.
For more highly-obscured stars, 
extinction becomes significant in the
near-IR, and $JHK_s$ photometry should provide more robust
 $A_V$ measurements.

\subsection{Systematics Due to the Choice of Model Atmospheres\label{sys}}
One potentially important source of systematic errors in this analysis
is the choice of stellar atmosphere models. The ATLAS9 atmospheres
\citep{Kurucz} were
chosen for convenience, as they 
come packaged with the \citet{fitter} fitting tool. These atmospheres
are completely static and
assume local thermodynamic equilibrium (LTE), therefore they are not strictly 
appropriate for hot stars, in particular hot stars with strong winds. 
In the {\it static} atmosphere case, spectral {\it shapes}
are not significantly affected by
the inclusion of non-LTE effects, and LTE atmospheres are
adequate for the analysis of low-resolution spectra or SEDs
\citep{TLUSTY}. 
However, O-star parameters computed using fully line-blanketed, non-LTE
{\it expanding} atmospheres show significant changes in the calibration of
$T_{\rm eff}$ and surface gravity ($\log{g}$) versus spectral type compared with
static models, and these effects are, not surprisingly, most
pronounced for stars with the strongest winds \citep{MSH05}. 
Surface gravities for the hottest atmospheres in the ATLAS9 model grid
range from $\log{g/[{\rm cm~s^{-2}}]}=4.0$ to 5.0, systematically high compared to the
\citet{MSH05} calibrations for O dwarfs ($\log{g/[{\rm cm~s}^{-2}]}=3.92$) and
especially O supergiants ($\log{g/[{\rm cm~s}^{-2}]}=3.2$ to 3.7).

To quantify the systematic errors introduced into the SED fitting
results by the adoption of the ATLAS9 atmospheres, we compared
publicly-available CMFGEN model spectra \citep{CMFGEN} to the nearest ATLAS9
spectra in $(T_{\rm eff},\log{g})$ parameter space. In our
comparisons, we normalized pairs of test
spectra to unity (equal $L_{\rm bol}$) and then computed the fractional
difference in flux density between the CMFGEN and ATLAS9 models in
each optical and IR bandpass used for SED fitting. A qualitative trend
readily emerged, in which the CMFGEN models compared to the ATLAS9
models predict similar flux 
density in $UBV$ but higher
flux density in the IR balanced by lower flux density in the
far-UV. This
redistribution of flux from the 
far-UV to the IR is attributable to free-free emission in the stellar
winds. Our SED fitting analysis, which effectively scales the ATLAS9 models
to the IR SED, may therefore 
{\it overestimate} $L_{\rm bol}$, compensating by overestimating
$A_V$. We define the magnitude of the systematic error predicted by these comparisons as
\begin{displaymath}
  \Delta L_{\rm bol}\equiv \frac{L_{\rm bol}({\rm ATLAS9})-L_{\rm
      bol}({\rm CMFGEN})}{L_{\rm bol}({\rm CMFGEN})},
\end{displaymath}
which varies strongly with stellar surface gravity. We evaluated four
``test cases'' based on CMFGEN models spanning the full range of
stellar properties among the validation sample and summarize the
results in Table~2. While $\Delta L_{\rm bol}\sim 20\%$ for the OB
supergiants, the majority of stars in the
validation sample are late O and early B dwarfs and giants with
$\Delta L_{\rm bol}\le 5\%$, and the mean for the entire validation
sample is $\overline{\Delta L}_{\rm bol}= 5\%$.


\begin{deluxetable}{@{\hspace{-2mm}}c@{\hspace{-2mm}}c@{\hspace{-2mm}}c@{\hspace{-2mm}}c@{\hspace{-2mm}}c@{\hspace{-3mm}}c@{\hspace{-3mm}}c}\setcounter{table}{2}
\tabletypesize{\scriptsize}
\tablecaption{ Expected Model-Based Systematic Errors in SED Fitting Results\label{syserr}}
\tablewidth{0pt}
\tablehead{
  \colhead{CMFGEN} & \colhead{} & \colhead{$T_{\rm eff}$} &
  \colhead{$\log{g}$} & \colhead{$\Delta L_{\rm bol}$} &
  \colhead{$\Delta \log{L_{\rm bol}}$} & \colhead{$\Delta A_V$} \\
  \colhead{Test Case} & \colhead{$N_{\star}$} & \colhead{(kK)} &
  \colhead{(cm~s$^{-2}$)} & \colhead{(\%)} & \colhead{(dex)} & \colhead{(mag)}
}
\startdata
O2--O6.5 I/II  & 1  & 42.5 & 3.75 & 28 & 0.12 & 0.26 \\
O7--B I/II     & 20 & 30.0 & 3.25 & 19 & 0.08 & 0.19 \\
O3--O6.5 V/III & 20 & 42.5 & 4.00 & 14 & 0.06 & 0.14 \\
\tableline
O7--O9.5 V/III & 39 & 30.0 & 4.00 &  5 & 0.02 & 0.05
\enddata
\tablecomments{$N_{\star}$ is the number of stars in the validation
  sample corresponding to each test case, based on their spectral
  types and luminosity class (G11). The systematic errors expected for the 102
  B stars (excluding supergiants) are negligibly small. The first
  three rows constitute the set of 41 low-$g$ stars highlighted in
  Figures~\ref{lbolvlbol}--\ref{avvav}.} 
\end{deluxetable}

%
\begin{figure}
\epsscale{1.0}
\plotone{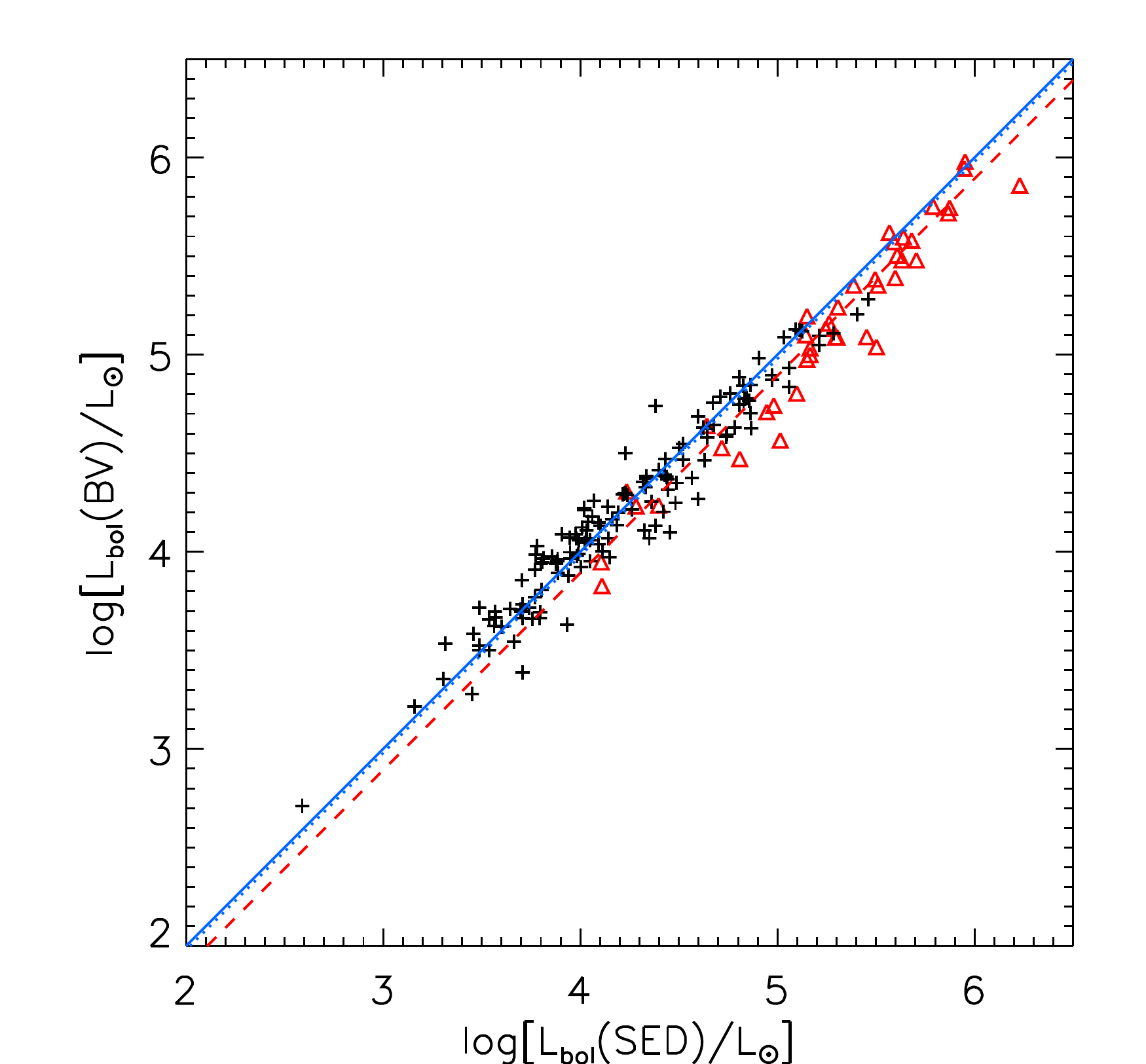}
\caption{Plot of $L_{\rm bol}(BV)$, derived through $BV$ color excess
  and bolometric correction (G11), versus $L_{\rm bol}({\rm SED})$ from
  optical+IR SED fitting for 181 stars in the validation sample
  (Table~1).  
  High-$g$ stars (late O and B dwarfs and giants) are plotted as
  crosses while low-$g$ stars (early O stars and OB supergiants) are
  plotted as (red) triangles. The dashed (red) line represents the
  horizontal offset from a 1--1 relation (solid
  blue line) expected from the {\it maximum} 28\% systematic error in $L_{\rm
    bol}({\rm SED})$, which {\it only}
  applies to the low-$g$ stars. The average (${\sim}5\%$) systematic
  offset for the entire validation sample is plotted as the dotted (blue)
  line. Luminosities assume an extinction law with
  $R_V=4$. The typical random uncertainty in $L_{\rm bol}(BV)$ is
  comparable to the symbol size, and the random uncertainty in $L_{\rm
    bol}({\rm SED})$ is less. 
\label{lbolvlbol}}
\end{figure}
%
In Figure~\ref{lbolvlbol}, we plot luminosity $\log{L_{\rm bol}(BV)}$
derived from $BV$ color 
excess and bolometric corrections based on CMFGEN models \citep[][G11]{MSH05}
against luminosity $\log{L_{\rm bol}({\rm SED})}$ from visible+IR SED
fitting with the ATLAS9 atmospheres, for 181 validation 
stars in Table~1.\footnote{Tr16-74 
lacks a  measurement in $B$ and was excluded.} Assuming typical random errors
in $UBV$ photometry of ${\sim}0.02$~mag \citep{MJ93}, the
random uncertainties on the luminosities plotted in
Figure~\ref{lbolvlbol} are smaller than the plotting symbol sizes.
There is generally
good agreement between the two methods, although deviations from the
1--1 relation that are greater than expected due to random photometric
errors are observed. In particular, the distribution of OB supergiants
and early O stars (the ``low-$g$'' stars; first 3 rows of
Table~2) is displaced toward higher values of $\log{L_{\rm 
    bol}({\rm SED})}$, as expected from the maximum model-based systematic
errors (dashed line in Figure~\ref{lbolvlbol}), but we stress
that the {\it average} systematic error (dotted line) not
significant. The scatter apparent in Figure~\ref{lbolvlbol} about the
1--1 relation among the late 
O and B dwarfs and giants (the ``high-$g$'' stars) is not due to our
choice of model atmospheres and instead reflects variations in
the extinction law, which impact $\log{L_{\rm bol}(BV)}$ far more
strongly than $\log{L_{\rm
    bol}({\rm SED})}$, as discussed in the following subsection (\S\ref{rv}).

Ideally, we would use the WM-basic \citep{WM-basic} or CMFGEN \citep{CMFGEN}
non-LTE, expanding atmosphere models for O stars. 
In a future upgrade to the SED-fitting method we will incorporate an
appropriate grid of models into the \citet{fitter} fitting tool, but
after accounting for the potential systematics, our
adoption of ATLAS9 atmospheres is not expected to introduce 
significant errors in the results of the present work. 
It is critical, however, to use the latest models to obtain
bolometric corrections as a function of spectral type, or for our
purposes, to assign the appropriate $T_{\rm eff}$(ST) to the stars in
the validation sample (Table~1).

\subsection{An ``Anomalous'' Extinction Law in the Carina
  Molecular Cloud\label{rv}}

\citet{WH76} suggested that the extinction law toward the
stellar population of the Carina Nebula was
best represented by a ratio of 
total-to-selective extinction $R_V=A_V/E(B-V)=5$, significantly higher
than the normal diffuse interstellar medium (ISM) value of
$R_V=3.1$. 
This claim of ``anomalous'' extinction was challenged by
\citet{TM80}, who measured $R_V=3.2\pm 0.28$,
consistent with the normal ISM value. 
\citet{MJ93} found $E(U-B)/E(B-V)=0.73\pm 0.01$ among their sample of
massive stars in Tr 16 and Tr 14, equivalent to $R_V=3.1$--3.2 and
yielding a spectroscopic distance modulus of $12.55\pm 0.08$ or
$d=3.2\pm0.1$~kpc, 
in significant disagreement with the now well-established distance of
$2.3$~kpc \citep{NS06b}. 
\citet{NW95} pointed out that adopting $R_V=4.0$ would lower the
spectroscopic distance modulus and bring it into agreement with the
geometric distance. Motivated by this historical debate, we performed
our SED fitting analysis on the validation sample twice, using
extinction laws characterized by $R_V=3.1$ and $R_V=4.0$, to evaluate
which choice better represents the {\it average} extinction law observed
toward the known massive stars in Carina.

Measurement of $A_V$ for a single star from $BV$ photometry, hereafter
denoted $A_V(BV)$,
depends on an accurate $M_V$, which in turn is based on knowledge of
$T_{\rm eff}$ and a 
bolometric correction (G11). As we demonstrated in \S\ref{lbolavsed} above,
fitting the full optical through IR SED separates the
problem, providing a measure of $L_{\rm bol}$ from the IR that is
generally independent of the optical photometry.
The result is an improved determination of $A_V$ that is also less
sensitive to the adopted extinction law.
If the actual extinction law is characterized by $R_V=4.0$, then
assuming $R_V=3.1$ will yield measurements of $A_V(BV)$ that are
systematically {\it underestimated} by 22.5\%. We find empirically from the
SED fitting results that $A_V$(SED) only increases by 6\%
when an extinction law characterized by $R_V=4.0$ is adopted in place of the
default extinction law characterized by  $R_V=3.1$ \citep{CCM89,I05}. 
Because the impact of the extinction law is different for the two
methods, 
the preferred average value of $R_V$ is the one that {\it maximizes} agreement
between $A_V({\rm SED})$ and $A_V(BV)$.

%
\begin{figure*}
\epsscale{1}
\plottwo{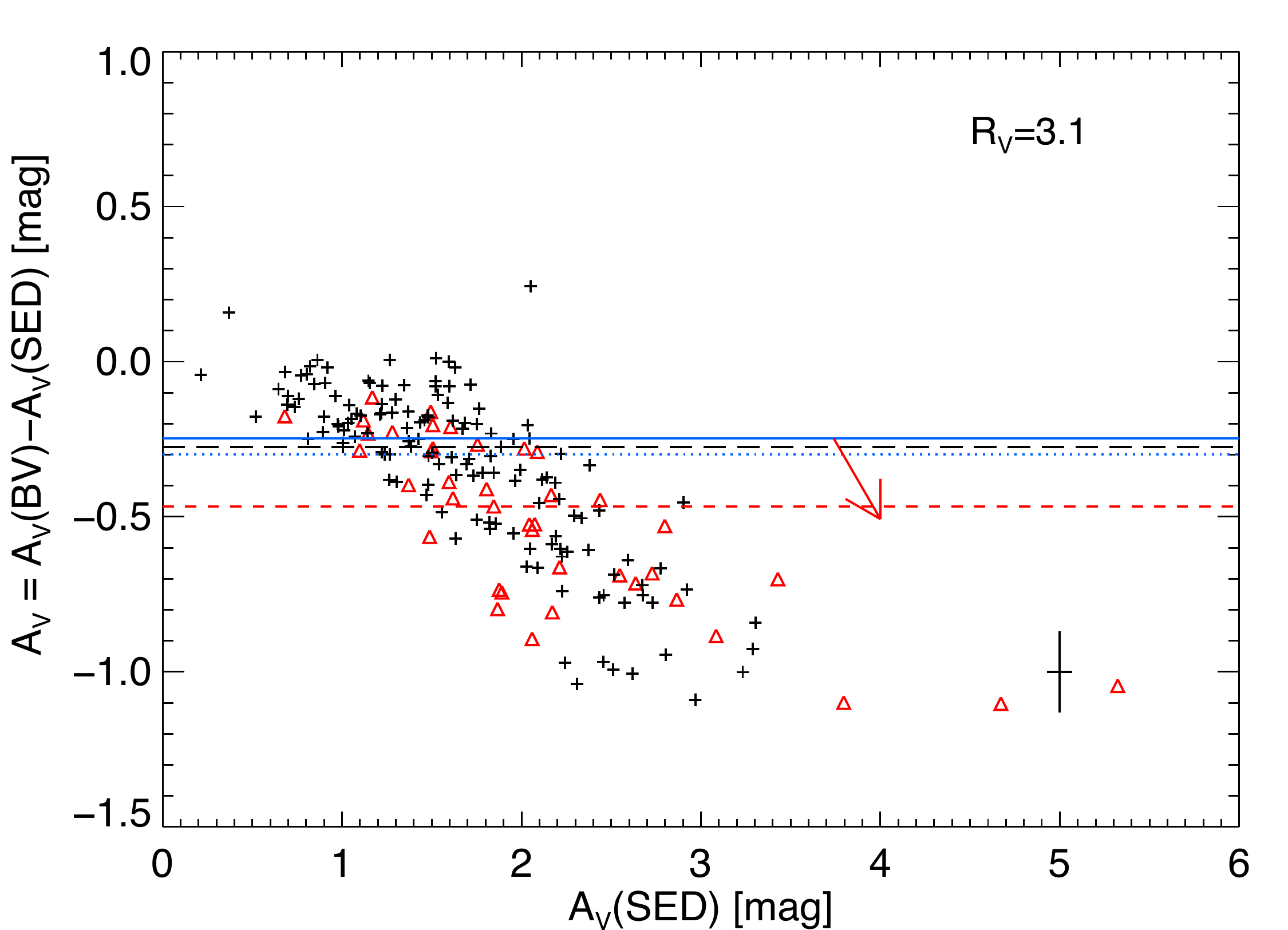}{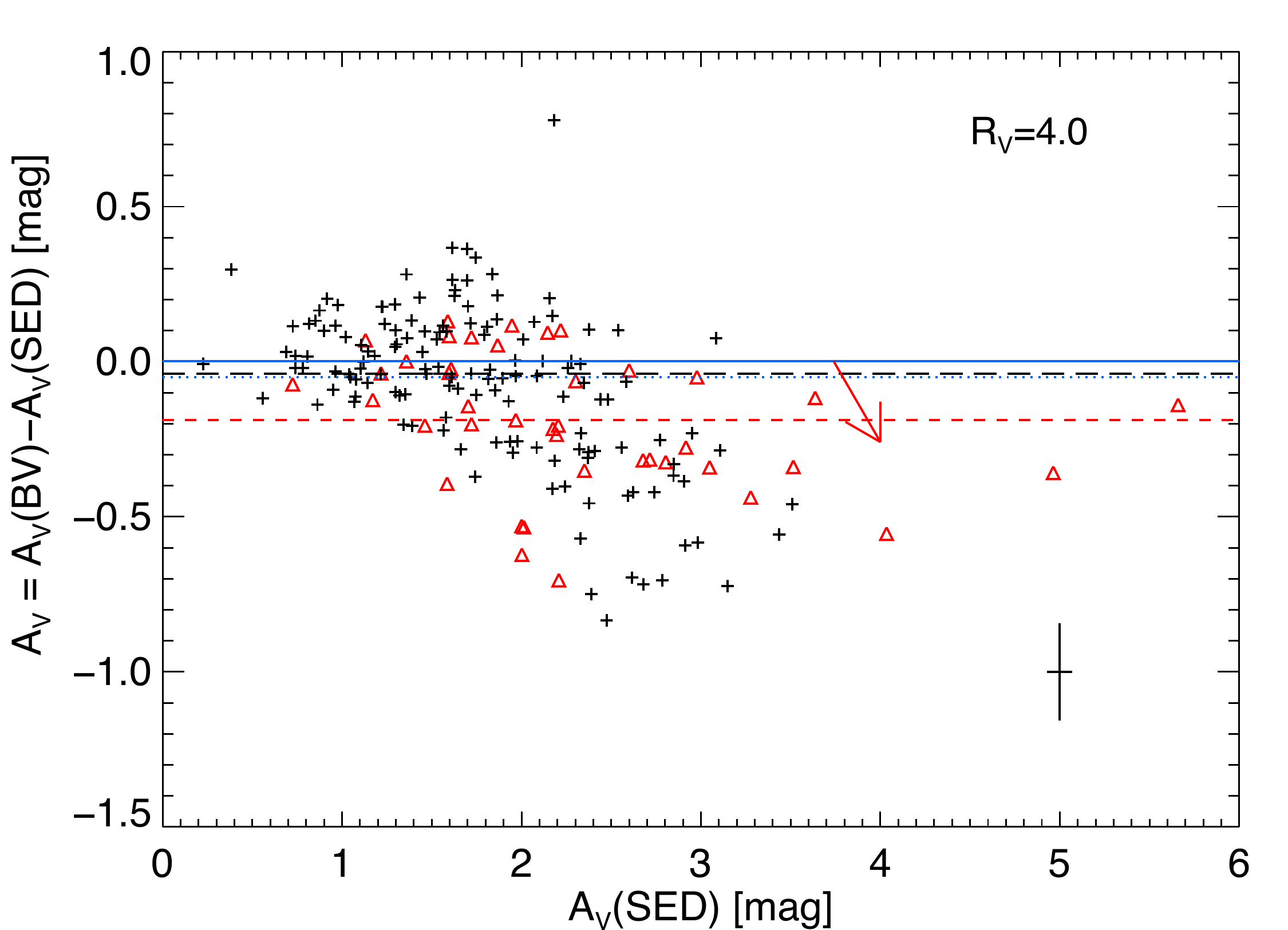}\\
\plottwo{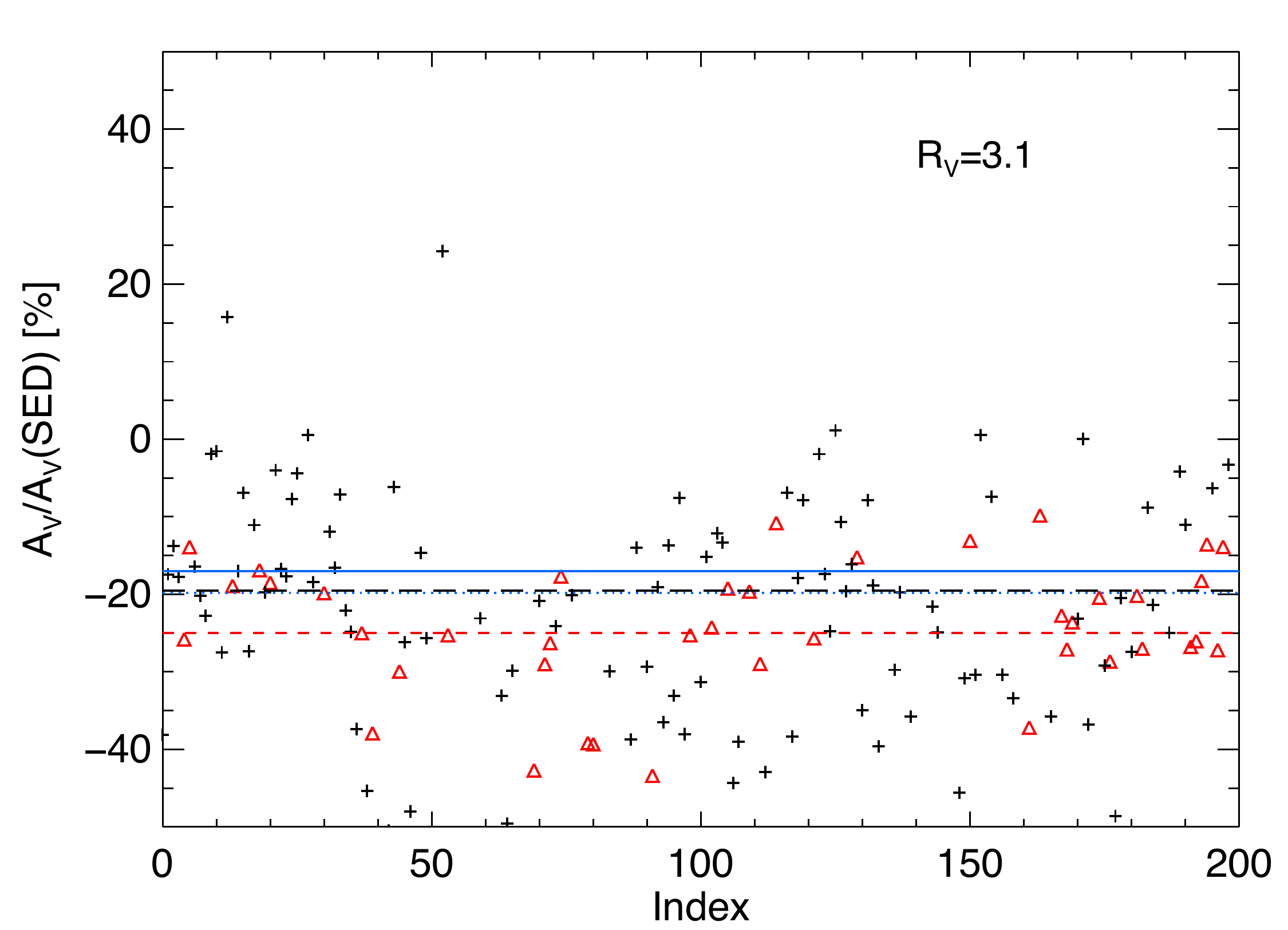}{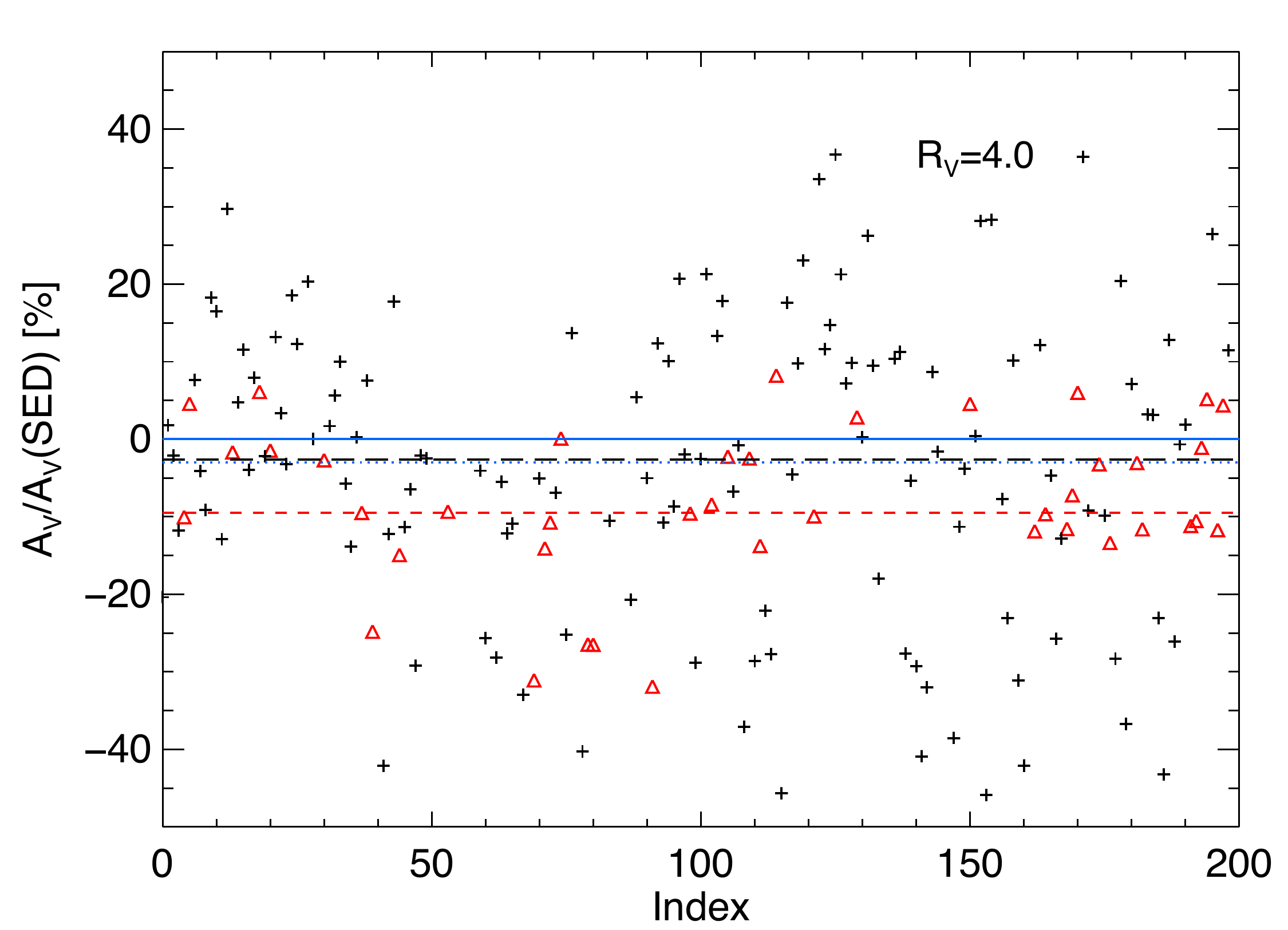}\\
\caption{Comparison between $A_V$ determined from optical+IR SED
  fitting and from
  $BV$ color excess (G11) for two different
  extinction laws, characterized by 
  $R_V=3.1$ and 4.0 (\S\ref{rv}). Plotting symbols are the same
  as in Figure~\ref{lbolvlbol}. Top panels: Absolute $\Delta A_V$ plotted against
  $A_V$ measured from SED fitting. The typical error bars due to
  random photometric uncertainties are shown. The (red) arrows
  indicate the {\it maximum} systematic error in the SED modeling;
  this applies {\it only} to the low-$g$ stars (red triangles).
  Bottom panels:
  Percentage $\Delta A_V/A_V$(SED) plotted against stellar index
  number (G11). In all panels, the long-dashed line is the
  median of the high-$g$ distribution, the dashed (red) line is the
  median of the low-$g$ distribution, the dotted (blue) line is the
  median of all stars and the solid (blue) line is the median of all
  stars corrected for the mean systematic error. 
\label{Avcomp}}
\end{figure*}
We computed $\Delta A_V \equiv A_V(BV) - A_V$(SED) for 181 stars in the
validation sample using both extinction laws,
$R_V=3.1$ and $R_V=4.0$ (Figure~\ref{Avcomp}). 
In the top-left panel, it is apparent that $\Delta A_V$ is consistent
with zero only for lightly-reddened stars with $A_V({\rm SED})\le
1.8$~mag, while there is a trend toward negative $\Delta A_V$ with
increasing $A_V({\rm SED})$. This suggests that $A_V(BV)$ is
systematically underestimated using $R_V=3.1$. We note that the effect
of the model-based systematic error in the SED fitting trends in a
similar direction (arrow), but this error applies {\it only} to the
41 low-$g$ stars, and indeed it can be seen that the apparent trend of the
low-$g$ stars is simply the high-$g$ trend
transformed by the expected systematics.
In the bottom-left panel, the {\it median} $\Delta A_V/A_V({\rm SED})
=-18\%$ for $R_V=3.1$,
corrected for the expected average systematic error (\S\ref{sys}), as
 expected if 
the actual average extinction law were better characterized by $R_V=4.0$.  
Adopting $R_V=4.0$ (Figure~\ref{Avcomp}, right
panels) brings the corrected median $\Delta A_V$ to zero. We therefore
conclude that while $R_V=3.1$ is appropriate for the least-obscured
stars in Carina, the {\it 
  average} extinction law for the validation
sample is better characterized by $R_V=4.0$.  


%
\begin{figure}
\epsscale{1.0}
\plotone{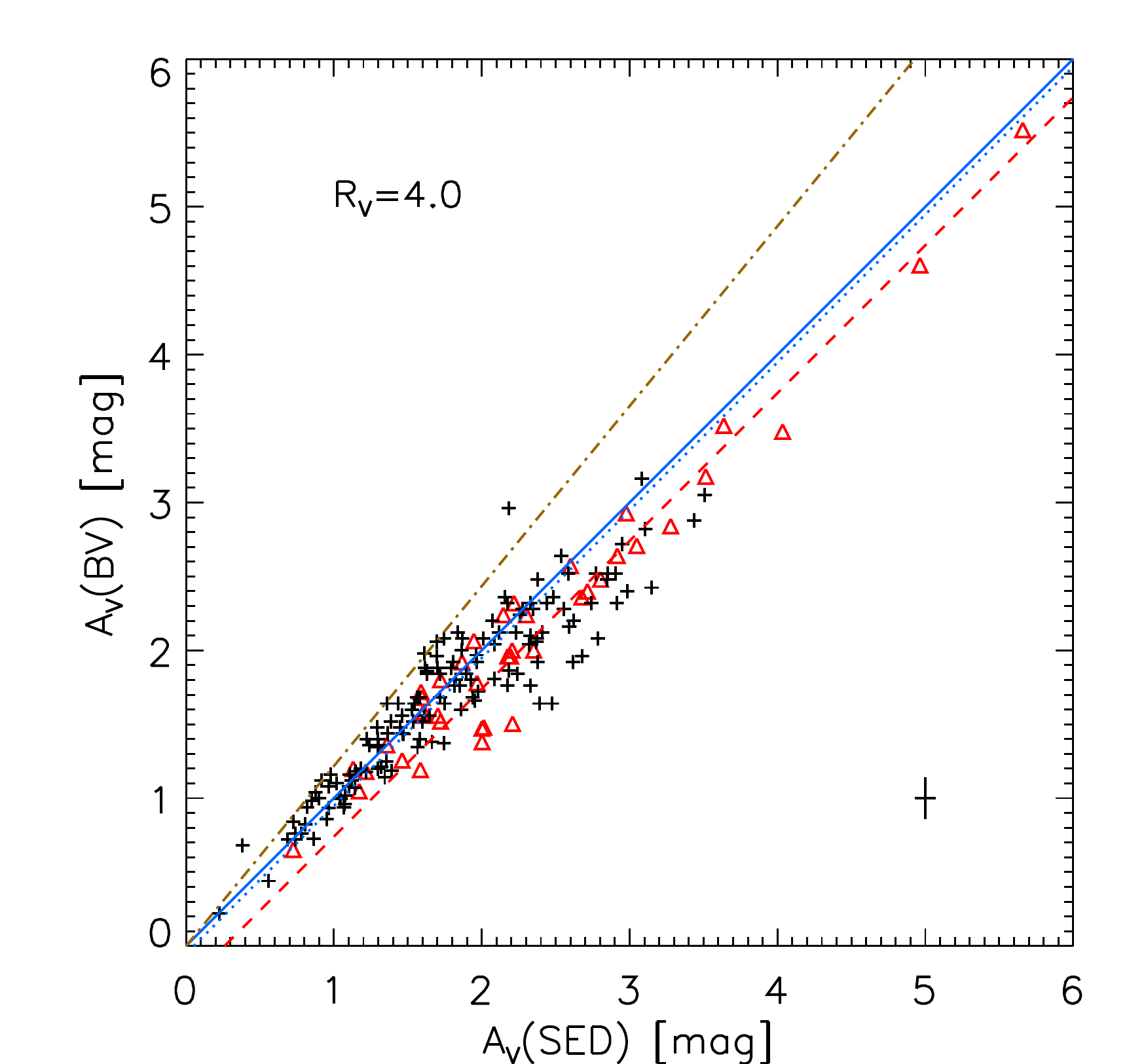}
\caption{Plot of $A_V(BV)$ versus $A_V({\rm SED})$
   for $R_V=4$ (Table~1). Plotting symbols and lines are the same as in
  Figure~\ref{lbolvlbol}, except a dash-dotted (brown) line has been
  added to show the transformation of the 1--1 line (solid blue) to
  the case of $R_V=3.1$. The typical error bars due to photometric 
  uncertainties are shown.
\label{avvav}}
\end{figure}
%
We subsequently adopt an extinction law characterized by $R_V=4.0$
\citep{CCM89}, since we are primarily interested in the global properties
of the massive stellar population rather than the detailed properties
of individual OB stars. The comparison of $A_V(BV)$ versus $A_V({\rm
  SED})$ using $R_V=4.0$ for 181 stars in the validation sample
(Table~1) is plotted in Figure~\ref{avvav}. As in
Figure~\ref{lbolvlbol}, the data 
agree with a 1--1 relation, although there is significant scatter due
both to random (photometric) uncertainties and systematic
effects. Also plotted is the
location of the 1--1 line transformed to the case of $R_V=3.1$,
a decrease of 22.5\% in $A_V(BV)$ and 6\% in $A_V({\rm SED})$.
Again there is a subset of stars with $A_V({\rm SED})\le
1.8$~mag, representing the upper envelope of the distribution, that agrees
better with $R_V=3.1$, but if we were to
adopt the standard ISM extinction law the large majority of sources
would fall {\it below} the 1--1 relation in the analogous plot to
Figure~\ref{avvav}. 
Among the validation sample (Table~1), we find
that $A_V$(SED) 
ranges from 0.2~mag to 5.7~mag, with mean $\overline{A}_V({\rm SED})= 1.9$~mag.

\section{Candidate X-ray-Emitting OB Stars}

The Vela--Carina Point-Source Catalog contains ${\sim}60,000$ sources within the
CCCP survey area (P11), dominated by stars
unassociated with the Carina complex. This would present an overwhelming
contamination problem  were we to attempt a search for candidate OB stars
using the near-IR and mid-IR data alone. 
The CCCP survey provides a crucial selection criterion. OB stars emit
primarily soft X-rays through
microshocks in their strong stellar winds \citep{FPP87},
although a variety of other mechanisms appear
to produce a heterogeneous mix of X-ray properties among the 200 known OB
stars in the Carina Nebula, 118 of which were detected by CCCP
\citep[G11;][]
{OBglobal}. 
Of the ${\sim}14,000$ sources in the CCCP X-ray
catalog,  ${>}10,000$ are likely young stars
in the Carina complex, and the vast majority of the IR counterparts to
CCCP sources are too faint 
to masquerade as OB stars at the Carina distance
\citep{CCCPcontam,CCCPclassifier}. 


\subsection{Sample Selection}

We began our search for
candidate OB stars with 3444 stars in the Vela--Carina Catalog that
were both well-fit by stellar atmosphere models and matched to CCCP
X-ray sources (Table~1 of P11).
We also examined 164 stars with marginal IR excess at 5.8 or 8 \um\
(P11) and CCCP counterparts that were well-fit by stellar
atmospheres when the band affected by excess emission was
excluded. Stars in the validation sample with CCCP detections were
included in these groups.
We then applied a magnitude cut at $K_s \le 14.6$, equivalent to a B3 V
star with $A_V=30$ mag extinction. This reduced the initial sample to
3183 stars without IR 
excess and 150 stars with marginal IR excess.
The primary purpose of this cut was to impose a requirement that the source
be detected in the 2MASS catalog; $JHK_S$ photometry is important for
constraining the SED models in the absence of optical photometry.

We then analyzed the loci of SED fitting results on the HR diagram for
each of the 3333 stars in 
the initial sample (as in Figure~\ref{HRD}). In contrast to the
validation sample analysis, we did not have any prior knowledge from 
which to constrain $T_{\rm eff}$, so we followed the procedure of \citet{WP08}
and identified the intersection point, if one existed, of the fit locus
with the theoretical OB MS, as defined by \citet{MSH05} for O
stars and \citet{dJN87} for B stars. As we would
expect, in the large majority 
(${\sim}90\%$) of cases, the fit loci fall below the OB MS plotted in
Figure~\ref{HRD}; these stars are insufficiently luminous, assuming
$d=2.3$~kpc, to be OB stars in Carina. For the remaining ${\sim}10\%$
of the initial sample, the intersection with the MS exists and is defined by
$T_{\rm eff}^{\rm MS}$. These stars could be OB stars in Carina,
but {\it only} if they have reached, or evolved beyond, the MS. It is apparent from
Figure~\ref{HRD} that a degeneracy between OB stars and lower-mass,
pre-main-sequence (PMS) stars exists where the locus of fits 
extends to cooler $T_{\rm eff}$ and crosses the PMS evolutionary
tracks \citep{SDF00}. 

For sufficiently luminous candidate OB stars, the degeneracy with PMS
stars can be broken. It is important to recall that, by design, {\it
  none} of the stars in our initial sample exhibit significant IR excess
emission, so if they are PMS stars, they can neither be embedded nor possess
optically thick circumstellar disks. This effectively places a lower
limit on the age of each star, corresponding to the destruction timescale for the
circumstellar disk and/or envelope. While this 
timescale is not well-known for intermediate-mass stars, it
likely decreases with increasing stellar mass \citep[][P11]{PW10}. Massive
stars reach the MS while still embedded \citep{ZY07}, 
hence we do not expect to find {\it any} massive
PMS stars in our sample.

\begin{figure*}
\epsscale{1.0}
\plotone{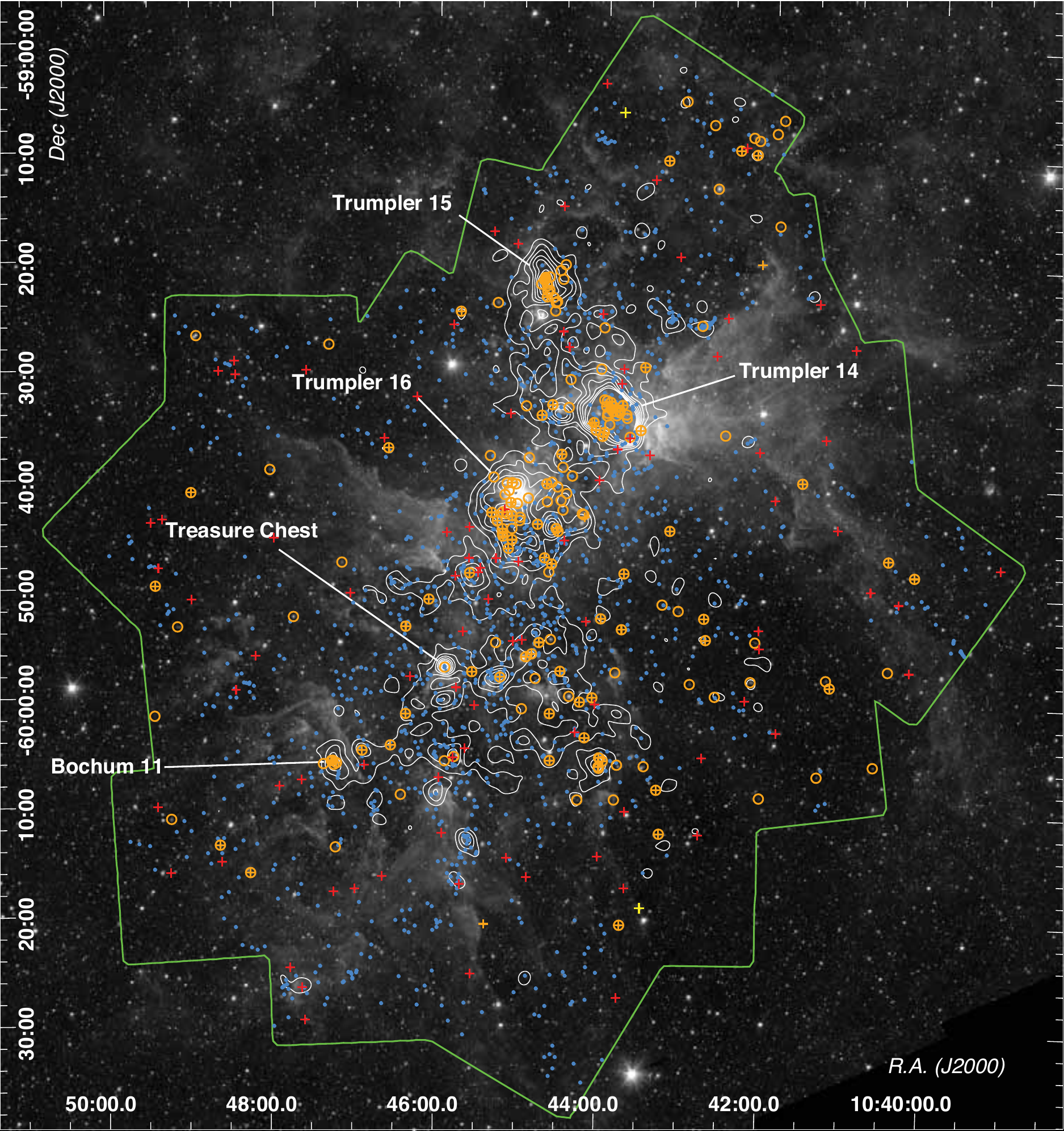}
\caption{{\it Spitzer} 3.6~\um\ image containing the full CCCP survey
  area (green outline). 
  Positions of OB stars in the validation sample are plotted as orange circles.
  The locations of X-ray-emitting OB stars previously confirmed by
  spectroscopy and identified by our SED
  fitting search criteria are overplotted as orange crosses (where
  such stars are also in the validation sample, the symbols become
  crossed-circles). The positions of the 94 candidate OB stars are shown
  as red crosses. Two foreground X-ray-emitting
  giant stars confirmed by
  spectroscopy are marked by yellow crosses. YSOs from P11 are plotted as small blue
  circles. Contours show surface density of CCCP sources 
  \citep{CCCPclusters}; prominent named star clusters are labeled.
\label{image}}
\end{figure*}

When the bolometric
luminosity of the MS intersection point falls above a fiducial
value of $\log{L_{\rm bol}^{\rm MS}/{\rm L}_{\sun}=4}$,
the SED fitting results become
physically inconsistent with a PMS interpretation. The three examples plotted
in Figure~\ref{HRD} illustrate the general cases of fit loci falling
below (blue), above (red), and near (green) the fiducial luminosity. 
Lacking an independent determination of $T_{\rm eff}$ from spectroscopy,
the (blue) stellar locus that intersects the MS near 
$(\log{T_{\rm eff}/{\rm K}},\log{L_{\rm
    bol}/{\rm L}_{\sun}})=(4.25,3)$ cannot be interpreted
unambiguously. Models to the left of the MS can be excluded, but
models to the right are consistent with PMS stars
between 0.5 and 2~Myr old.
Such a source could be a diskless,
X-ray-emitting, low- or intermediate-mass PMS star 
in the Carina Nebula, and this interpretation would be
preferred given that late B stars are not expected to emit X-rays
\citep{NE11}. In contrast, the (red)  
stellar locus with MS intersection near $(\log{T_{\rm eff}/{\rm K}},\log{L_{\rm
    bol}/{\rm L}_{\sun}})=(4.55,5)$ is unambiguously massive. The PMS
interpretation is effectively eliminated for this star, because 
it implies an extremely young age (${\ll}0.5$~Myr) that is
inconsistent with the lack of a strong 
IR excess.
Even if this star has evolved past the MS, we would not disqualify it for inclusion in our
list of candidate OB stars, because as a post-MS Carina member it would necessarily be
massive, most probably an OB giant or supergiant with $T_{\rm eff}<T^{\rm MS}_{\rm eff}$. 
Finally, the
(green) stellar
locus with MS intersection near $(\log{T_{\rm eff}/{\rm K}},\log{L_{\rm
    bol}/{\rm L}_{\sun}})=(4.4,4)$ represents the borderline case. The
PMS interpretation requires a diskless, X-ray-emitting star younger
than 0.5~Myr. Such an 
interpretation is possibly allowed if the disk lifetimes of
intermediate-mass stars are very short \citep{PW10}, but the OB
interpretation is more plausible and hence preferred. We therefore
find that stars with equivalent MS luminosity $\log{L_{\rm
    bol}^{\rm MS}/{\rm L}_{\sun}}\ge 4$ rise above the PMS degeneracy. Applying
this cut to our sample, we selected 179 candidate
X-ray-emitting OB stars 
(including 6 with marginal IR excess in the IRAC bands), shown on a
\spitzer\ image of the Carina Nebula in Figure~\ref{image}.
%

The selection procedure described above recovered 84 of the 182
sources in the validation sample. {\it All}
validation stars were {\it not} selected because: (1) nearly half of the 200
known OB stars in Carina lack CCCP detections (G11), (2) some
validation stars lack 2MASS photometry (Table~1), and (3)
validation stars with spectral types later than B1~V (Cl* Trumpler 16 MJ 323 in Figures
\ref{SEDs} and \ref{HRD}, for example) generally do not reach
the luminosity cutoff of $L_{\rm bol}^{\rm MS}\ge 10^4$~\Lsun\ for
unambiguous candidate OB stars.

As a final step in establishing our list of candidate OB stars in
Carina, we discarded sources with cataloged spectral types listed in
the \citet{Skiff09} compilation, 
including the 84 OB stars the validation
sample that have counterparts in the CCCP catalog and two foreground
giants that were also
selected by our search criteria. 
The final sample of 94 candidate X-ray-emitting OB stars (including 2 with
marginal IR excess) is presented in Table~3, and each star is
given a catalog number, preceded by ``OBc'', in column
(1).\footnote{OBc 94 falls just below the fiducial luminosity cutoff,
  but it was added to the sample because its high-quality X-ray
  spectrum strongly suggests a 
  massive star (see \S\ref{xprops}).} 
Values of $T_{\rm eff}^{\rm MS}$, $\log{L_{\rm bol}^{\rm MS}}$, and $A_V^{\rm MS}$ (using $R_V=4.0$) computed 
from the MS intersection points (Figure~\ref{HRD}) are listed in
columns (3)--(5). We
stress that these stellar parameters should only be used for comparison
purposes among the sample, because we have {\it assumed} that each
star is on the theoretical MS. Formally, these values 
are upper limits, given the physical constraint that $T_{\rm eff}\le
T_{\rm eff}$(MS). Systematic errors due to the adoption of the ATLAS9
stellar atmosphere models (see \S\ref{sys}) may also lead formally to
overestimates of $\log{L_{\rm bol}^{\rm MS}}$, and $A_V^{\rm MS}$, but
this effect is unimportant given that our analysis produces upper
limits anyway.
Unresolved binaries and evolved, post-MS OB stars
exist in the validation sample (G11) and
are expected to be present in the candidate OB sample as well.  
The parameters given
in Table~3 for such stars are certainly overestimates. The
OB classification of all stars, as well as the physical parameters
derived from SED fitting, must be regarded as tentative until
confirmed by follow-up spectroscopic observations.

We have checked for previous identifications of each star in Table 2
in existing stellar catalogs. Most, 
but not all, of the candidate OB stars were detected by
optical all-sky surveys listed in VizieR \citep{VizieR},
but the catalogs were intended
primarily for astrometry, 
and the photometry is generally of insufficient quality to aid in our SED
analysis. A few have previous identifications and tentative 
spectral classifications, listed in column (14).
Eleven of the candidate OB stars are cataloged
in the Henry Draper Extension Charts \citep[HDEC;][]{HDEC}, with the
following distribution of spectral classes (no luminosity classes are
given): 3 B-type, 5 A-type, 2 G-type, and 1 K0. 
More recent, optical/near-IR photometric observations, sometimes combined with
X-ray observations \citep{C93,NE03,S07}, identified 4 of our 
candidate OB stars and misclassified 2 others as candidate
PMS stars (one of the latter, OBc 42, is among the most highly-obscured
stars in our sample).

\subsection{Possible Contaminating Sources\label{contam}}

We have assumed that X-ray emission is a strong indicator that a
candidate OB star in the CCCP field
is located at the distance of
the Carina Nebula. \citet{CCCPcontam} provide Monte Carlo simulations that
evaluate the potential contaminating source populations expected in
the CCCP field. Background active galactic nuclei are the most numerous X-ray
contaminants, but such objects are far too faint in the IR to be mistaken
for OB stars. 
Galactic field stars, however, can be sufficiently bright in X-rays to have
been included in the CCCP catalog \citep{CCCPcontam}. In this section,
we evaluate the possible levels of contamination in our candidate OB
star sample from unassociated field stars.

{\it Foreground MS stars.}
\citet{CCCPcontam} predict ${\sim}1600$ foreground main-sequence stars
in the CCCP, broken 
down by spectral class into 150, 450, 140, and 850 F, G, K, and M
dwarfs, respectively. K and M dwarfs are too cool and faint to
be mistaken for OB stars by our SED analysis, but F and G
dwarfs could produce contamination, if they are not too distant. We
have estimated the   
maximum distance and potential numbers of foreground MS contaminants 
as follows. 
We iterated our analysis of the SED fitting results, experimenting
with assumed distances from $d=0.1$ to 0.5 kpc, and measured the
$L_{\rm bol}$ values for $T_{\rm eff}=5700$.
For $d\le 0.2$~kpc, a population of possible G dwarfs with $L_{\rm bol}\sim
1$~\Lsun\ emerged. 

The least-obscured stars in the validation sample
have $A_V\sim 1$~mag and
reddening consistent
with a diffuse ISM extinction law characterized by $R_V=3.1$
(Figures~\ref{Avcomp} \& \ref{avvav}). The 
extinction to the least-obscured 
Carina members is therefore dominated by the ISM along the
line-of-sight to the Carina 
Nebula with a negligible contribution from local cloud material.
The extinction to a nearby star should be a small fraction of the
${\sim}1$~mag extinction to the Carina complex, hence
we identified 17 potential foreground dwarfs satisfying $A_V< 0.09~{\rm
  mag}= (0.2~{\rm kpc}/2.3~{\rm kpc})\times 1$~mag, assuming $T_{\rm
  eff}=5700$ and a uniform distribution of extinction with distance.
The simulations predict ${<}20$
F and G MS stars at $d<0.2$~kpc detected by the CCCP \citep{CCCPcontam}.
The 17 candidate OB stars that could possibly be foreground
dwarfs are identified by ``PFD'' in column (14) of Table~3.



{\it Giants}. The simulations of \citet{CCCPcontam} predict 
${\la}400$ (150 foreground + 250 background) contaminating giant
stars of all spectral classes. Because giants, especially red giants
with K and M spectral types, are very luminous in the IR, they
represent a potentially serious source of contamination if they are
detected in significant numbers by the CCCP. Taking the simulations at
face value, we might expect to misclassify 400 X-ray-emitting giants
as candidate OB stars, distributed evenly throughout the CCCP
field. This is clearly not the case, which suggests that the actual
number of giants masquerading as OB stars is far lower. Late K and M giants
are sufficiently cool that they can be separated from hot stars on the
basis of their near-IR colors, hence they are unlikely to be mistaken
for OB stars in our SED analysis.
Furthermore, the giants most likely to be
strong X-ray emitters detected by CCCP are
F and G giants with ages ${<}1$~Gyr \citep{RR95,PMS00}. 
Such stars are relatively uncommon. The \citet{CCCPcontam}
simulations predict only ${\sim}20$ (10 foreground + 10 background) F
and G giants in
the CCCP catalog, and depending on the distance and extinction, these may
or may not be selected as candidate OB stars.


Based on the arguments above, the {\it maximum} contamination fraction among
our candidate OB stars is 
${\la}40\%$, and this could be a high upper limit.
We know of only a few unassociated stars chosen by our selection
criteria. Two spectroscopically identified giants (Figure~\ref{image})
were discarded, 
and an additional 8 stars 
from the HDEC \citep{HDEC} 
have non-OB spectral types listed, 6 of which we independently
flagged as possible foreground dwarfs (Table~3, column 14). 
One star, HD 305547 (OBc 72), has cataloged spectral type
K0 (Table~3) and probably is a giant, since this classification seems more
likely than a highly-luminous O star (equivalent to an O3 V star
on the MS) located far from the major
ionizing clusters and active star forming regions in the Carina Nebula.
It is not clear, however, that
the HDEC spectral types are reliable, since the candidate OB stars
fall at the extreme faint limit of the HDEC.
Considering its source crowding and contamination from bright nebular
emission,  the Carina Nebula 
posed an extremely
challenging target for the technology available for optical
spectroscopy in the early 20th century. 
New, targeted spectroscopy is needed to confirm the identifications of
all 94 candidate OB stars.

\subsection{Spatial Distribution}

The spatial distributions of known and candidate OB stars selected via our SED
fitting method  shows good correspondence with the
distribution 
of YSOs selected via mid-IR excess emission by P11 (Figure~\ref{image}).
OB stars, in particular the known OB stars (G11), are spatially
correlated with
the stellar clusters identified 
by \citet{CCCPclusters}, but there is 
clearly a more distributed population as well. Unsurprisingly, the
candidate OB stars are
preferentially found outside of the well-studied, lightly-obscured
ionizing clusters Tr 16, 14, 15 and Bo 11. The Treasure Chest cluster
contains 
young OB stars, but it is too compact and too contaminated by
nebulosity for reliable
2MASS and {\it Spitzer} point-source detections.

The average extinction ($R_V=4.0$) of the candidate OB stars is
$\overline{A}_V^{\rm MS}=
5.8$~mag. The minimum $A_V^{\rm MS} = 1.6$~mag, and it should be
noted that the least-obscured candidate OB stars, with $A_V^{\rm MS}<3$~mag
could instead be
contaminating foreground stars (Table~3).
The star in our sample with the highest extinction is OBc~59, with $A_V^{\rm
  MS}=35.6$~mag; this star is apparently located inside or behind an obscuring
IR dark cloud in the
dense molecular cloud known as the giant pillar \citep{NS00,JR04,YY05}.
OBc~59 is found in the middle of a small but significant
overdensity of CCCP sources \citep{CCCPclusters}; it is unlikely to be
an unassociated background star and may be the most massive member of
a highly-obscured young cluster. 
A similarly highly-obscured star is OBc 42, with $A_V^{\rm
  MS}=33.9$~mag; this star is apparently located near IRAS 
10430--5931, the first embedded cluster discovered in the Carina
Nebula \citep{TM96}.

Extinction is generally higher in the South Pillars
\citep{NS00,spitzcar} region, located southeast of Tr 16
(Figure~\ref{image}), and behind the 
V-shaped, obscuring dust lane south of Tr 16 and 14 \citep{SB07}. 
The greatest concentration
of candidate OB stars is found in the dust lane southeast of Tr 16, where
\citet{S07} and \citet{CCCPclusters} 
have found evidence for an obscured, massive cluster possibly
associated with a dense molecular core \citep{YY05}. The six
candidate OB stars in this group (OBc 48, 50, 51, 52, 56, and 61) have
average $\overline{A}_V^{\rm MS}=8.6$~mag, and three of them (OBc 50, 52, and
56) have $\log{L_{\rm bol}^{\rm MS}/L_{\sun}}\ge 5.6$ (Table~3),
  equivalent to O4 V stars \citep{MSH05}.


\subsection{X-ray Properties\label{xprops}} 





Among the candidate X-ray-emitting OB stars, 21 had X-ray spectra of
sufficiently high signal-to-noise (S/N) to measure the absorbing
column density
 $N_{\rm H}$, spectral temperature $kT$, and total-band (0.5--8~keV),
absorption-corrected X-ray luminosity 
$L_{t,c}$ by minimizing the $C$ statistic \citep[$C_{\rm stat}$;][]{Cstat}
implemented in the XSPEC spectral fitting package \citep{XSPEC}.
We define high-S/N sources as those whose spectra in the fit range of
0.5--8 keV could be divided 
into 8 or more groups (spectral bins) with S/N~$\ge 3$ per group. This
is similar to the requirement used by \citet{OBglobal} that a source
modeled with XSPEC have ${\ge}50$ net 
counts.
OBc~94 is a member of this high-S/N group with 511 total-band net
counts, a soft X-ray spectrum ($kT=0.47$~keV), and no significant variability, all of which
suggest a massive star, hence we included OBc~94 in our candidate OB
star sample even though its IR SED alone is not {\it quite} 
bright enough ($\log{L_{\rm bol}^{\rm MS}/L_{\sun}}=3.9$,
Table~3) to qualify as an unambiguous OB candidate.

We performed XSPEC fits to the spectra of the 21 high-S/N sources in
two different ways. (1) Similar to the approach of \citet{OBglobal},
we fit one- or two-component thermal plasma models, freezing the
absorption parameter to the value corresponding to $A_V^{\rm MS}$
found by the IR SED fitting,
$N_{\rm H}^{\rm MS}$ \citep[assuming $N_{\rm H}/A_V=1.6\times
10^{21}$~cm$^{-2}$ mag$^{-1}$;][]{MV03}. The X-ray properties returned
by these ``frozen-absorption'' fits are presented in Table~4. But
whereas \citet{OBglobal} analyzed {\it known} Carina OB
stars for which the measurement of $A_V$ is
robust,\footnote{\citet{OBglobal} used $A_V($SED) from our
  Table~1, if available, and $A_V(BV)$ from G11 otherwise.}
until $T_{\rm eff}$ can be measured by follow-up optical/near-IR
spectroscopy $N_{\rm H}^{\rm MS}$ is simply an upper limit to the 
absorption, and  we cannot even be
certain that a given candidate OB star is actually a Carina member. To
obtain a measure of $N_{\rm H}$ that is independent 
of the assumptions used in the IR SED fitting, (2) we again fit the X-ray
spectra with one- or two-component thermal plasma models, this time letting
$N_{\rm H}$ vary as a free parameter. The X-ray properties 
returned by these ``free-absorption'' fits are presented in Table~5.
\begin{deluxetable*}{rcrcllcc}\setcounter{table}{4}
\tablecaption{X-ray Spectral Fitting Results for 21 Candidate OB Stars
  with High S/N Spectra, Absorption Frozen to $A_V^{\rm MS}$\label{frozen}}
\tablewidth{0pt}
\tablehead{
\colhead{(1)} & \colhead{(2)} & \colhead{(3)} & \colhead{(4)} & \colhead{(5)} & \colhead{(6)} & \colhead{(7)\tablenotemark{b}} & \colhead{(8)\tablenotemark{b}} \\
\colhead{OBc} & \colhead{} & \colhead{Net Counts} & \colhead{$\log{N_{\rm H}^{\rm MS}}$} & \colhead{$kT_1$} & \colhead{$kT_2$} & \colhead{$\log{L_{t,c,1}}$} & \colhead{$\log{L_{t,c,2}}$} \\
\colhead{No.} & \colhead{CXOGNC J\tablenotemark{a}}  & \colhead{0.5--8~keV} & \colhead{(cm$^{-2}$)} & \colhead{(keV)} & \colhead{(keV)} & \colhead{(erg~s$^{-1}$)}  & \colhead{(erg~s$^{-1}$)}
}
\startdata
  1 &   103909.94-594714.5      & 1864 & $21.8\phd$ & ${\phn}0.20\phd_{-0.01}^{+0.03}$  & $1.0\phd_{\cdots}^{+0.1}$ & 33.12 & 31.98 \\ 
  2 &   104014.67-595654.4    &  210 & $21.7\phd$ & ${\phn}0.56\phd_{-0.08}^{+0.06}$  & \nodata                   & 31.72 &   \nodata \\ 
  5 &   104059.29-592724.9     &  146 & $21.9\phd$ & ${\phn}0.86\phd_{-0.2}^{+0.2}$    & $1.7\phd_{\cdots}^{+1.2}$ & 31.35 & 30.86 \\ 
 10 &   104154.91-594123.6      & 1249 & $21.8\phd$ & ${\phn}0.20\phd_{-0.02}^{+0.05}$  & $1.6\phd_{-0.10}^{+0.10}$ & 32.36 & 31.67 \\ 
 12 &   104205.01-595317.4     &  146 & $21.7\phd$ & ${\phn}0.38\phd_{-0.06}^{+0.09}$  & \nodata                   & 31.75 &   \nodata \\ 
 18 &   104246.53-601207.0    & 1403 & $21.6\phd$ & ${\phn}0.34\phd_{-0.05}^{+0.04}$  & $1.8\phd_{-0.3}^{+0.3}$   & 32.37 & 31.78 \\ 
 19 &   104306.95-591915.0    &  139 & $22.1\phd$ & ${\phn}2.1\phd_{-0.4}^{+0.7}$     & \nodata                   & 31.30 &   \nodata \\ 
 30 &   104401.63-590327.4     &  174 & $21.7\phd$ & ${\phn}0.22\phd_{-0.05}^{+0.06}$  & $1.5\phd_{-0.3}^{+0.4}$   & 31.77 & 31.00 \\ 
 32 &   104402.75-593946.0  & 1278 & $21.9\phd$ & ${\phn}0.18\phd_{-0.04}^{+0.07}$  & $1.6\phd_{-0.09}^{+0.11}$ & 32.37 & 31.64 \\ 
 34 &   104411.16-595242.6     &  198 & $21.6\phd$ & ${\phn}0.27\phd_{-0.03}^{+0.03}$  & \nodata                   & 31.93 &   \nodata \\ 
 39 &   104430.89-591446.0     &  201 & $21.9\phd$ & ${\phn}3.0\phd_{-0.6}^{+1.1}$     & \nodata                   & 31.70 &   \nodata \\ 
 41 &   104457.51-595429.5     &  574 & $21.8\phd$ & ${\phn}0.23\phd_{-0.09}^{+0.07}$  & $1.4\phd_{\cdots}^{+0.2}$ & 32.06 & 31.50 \\ 
 50 &   104522.29-595047.0     &  379 & $21.9\phd$ & ${\phn}0.61\phd_{-0.2}^{+0.2}$    & $1.6\phd_{-0.4}^{+1.4}$   & 31.46 & 30.81 \\ 
 52 &   104530.22-594821.0     &  218 & $22.2\phd$ & ${\phn}2.4\phd_{-0.5}^{+0.7}$     & \nodata                   & 31.49 &   \nodata \\ 
 55 &   104536.45-594410.7 &  222 & $22.1\phd$ & ${\phn}8.1\phd_{-2.9}^{\cdots}$   & \nodata                   & 31.61 &   \nodata \\ 
 56 &   104536.75-594702.2 &  220 & $22.1\phd$ & ${\phn}0.55\phd_{-0.1}^{+0.2}$    & $1.4\phd_{-0.2}^{+0.5}$   & 31.73 & 30.95 \\ 
 57 &   104538.70-600426.5     &  150 & $21.6\phd$ & ${\phn}2.3\phd_{-0.5}^{+0.9}$     & \nodata                   & 31.08 &   \nodata \\ 
 67 &   104615.19-593217.6     &  133 & $21.6\phd$ & ${\phn}0.32\phd_{-0.1}^{+0.3}$    & $1.8\phd_{-0.4}^{+0.8}$   & 31.16 & 30.86 \\ 
 75 &   104735.26-602923.4    &  199 & $21.6\phd$ & ${\phn}0.52\phd_{-0.1}^{+0.1}$    & $1.6\phd_{-0.3}^{+0.5}$   & 31.29 & 31.12 \\ 
 88 &   104858.62-595057.4     &  288 & $21.6\phd$ & ${\phn}0.35\phd_{\cdots}^{+0.1}$  & $2.2\phd$                 & 31.59 & 30.15 \\ 
\hline
 94 &   104220.83-590908.6 & 511 & $21.6\phd$ & ${\phn}0.47$ & \nodata & 31.63 & \nodata 
\enddata
\tablenotetext{a}{CCCP Catalog identifier \citep{CCCPcatalog}.}
\tablenotetext{b}{X-ray luminosity in each thermal plasma component, corrected for the absorption given in column (4).}
\end{deluxetable*}


\begin{deluxetable*}{rcrl@{}l@{~}l@{}c@{}c@{}rrr}
\tablecaption{X-ray Spectral Fitting Results for 21 Candidate OB Stars
  with High S/N Spectra, Absorption As Free Parameter\label{free}}
\tablewidth{0pt}
\tablehead{
  \colhead{(1)} & \colhead{(2)} & \colhead{(3)} & \colhead{(4)} & \colhead{(5)} & \colhead{(6)} & \colhead{(7)} & \colhead{(8)} & \colhead{(9)} & \colhead{(10)} & \colhead{(11)} \\
  \colhead{OBc} & \colhead{} & \colhead{Net Counts} & \colhead{$\log{N_{\rm H}^{\rm Xfit}}$} & \colhead{$kT_1$} & \colhead{$kT_2$} & \colhead{$\log{L_{t,c,1}}$} & \colhead{$\log{L_{t,c,2}}$} & \colhead{$A_V^{\rm Xfit}$} & \colhead{$A_V^{\rm MS}-A_V^{\rm Xfit}$} & \colhead{} \\
  \colhead{No.} & \colhead{CXOGNC J} & \colhead{0.5--8~keV} & \colhead{(cm$^{-2}$)} & \colhead{(keV)} & \colhead{(keV)} & \colhead{(erg~s$^{-1}$)} & \colhead{(erg~s$^{-1}$)} & \colhead{(mag)} & \colhead{(mag)} & \colhead{$\Delta C_{\rm stat}$}
}
\startdata
  1 &   103909.94-594714.5 & 1864 & $21.5\phd_{-0.2}^{+0.1}$   & ${\phn}0.24\phd_{-0.02}^{+0.03}$ & ${\phn}1.0\phd_{\cdots}^{+0.2}$ & 32.56 & 32.04 &  2.1 &   1.5 &  11.6 \\ 
  2 &   104014.67-595654.4 &  210 & $21.7\phd_{-0.2}^{+0.1}$   & ${\phn}0.55\phd_{-0.15}^{+0.08}$ & \nodata                         & 31.78 &   \nodata &  3.5 &  -0.5 &   0.5 \\ 
  5 &   104059.29-592724.9 &  146 & $22.3\phd_{-0.08}^{+0.07}$ & ${\phn}0.68\phd_{-0.1}^{+0.1}$   & \nodata                         & 32.16 &   \nodata & 11.1 &  -6.3 &  25.0 \\ 
 10 &   104154.91-594123.6 & 1249 & $20.9\phd_{\cdots}^{+0.3}$ & ${\phn}0.78\phd_{-0.1}^{+0.2}$   & ${\phn}2.2\phd_{-0.2}^{+0.3}$   & 31.01 & 31.75 &  0.5 &   3.5 &  11.5 \\ 
 12 &   104205.01-595317.4 &  146 & $21.6\phd_{-0.6}^{+0.2}$   & ${\phn}0.45\phd_{-0.1}^{+0.2}$   & \nodata                         & 31.55 &   \nodata &  2.3 &   1.1 &   1.5 \\ 
 18 &   104246.53-601207.0 & 1403 & $20.0\phd_{\cdots}^{+1.0}$ & ${\phn}0.74\phd_{-0.11}^{+0.05}$ & ${\phn}2.1\phd_{-0.2}^{+0.5}$   & 31.78 & 31.86 &  0.1 &   2.6 &  17.6 \\ 
 19 &   104306.95-591915.0 &  139 & $22.3\phd_{-0.3}^{+0.2}$   & ${\phn}1.6\phd_{-0.7}^{+1.4}$    & \nodata                         & 31.41 &   \nodata & 11.3 &  -2.9 &   0.5 \\ 
 30 &   104401.63-590327.4 &  174 & $21.8\phd_{\cdots}^{+0.2}$ & ${\phn}0.19\phd_{-0.06}^{+0.10}$ & ${\phn}1.4\phd_{-0.2}^{+0.2}$   & 32.20 & 30.96 &  4.3 &  -1.2 &   0.3 \\ 
 32 &   104402.75-593946.0 & 1278 & $21.2\phd_{-0.3}^{+0.2}$   & ${\phn}1.1\phd_{-0.3}^{+0.5}$    & ${\phn}2.6\phd_{-0.4}^{+1.2}$   & 30.96 & 31.69 &  0.9 &   3.9 &  17.4 \\ 
 34 &   104411.16-595242.6 &  198 & $18.3\phd_{\cdots}^{\cdots}$                 & ${\phn}0.46\phd_{-0.09}^{+0.09}$ & \nodata                         & 31.31 &   \nodata &  0.0 &   2.6 &   9.7 \\ 
 39 &   104430.89-591446.0 &  201 & $22.0\phd_{-0.2}^{+0.2}$   & ${\phn}2.6\phd_{-0.8}^{+1.3}$    & \nodata                         & 31.74 &   \nodata &  6.9 &  -1.5 &   1.2 \\ 
 41 &   104457.51-595429.5 &  574 & $20.4\phd_{\cdots}^{+0.8}$ & ${\phn}0.86\phd_{-0.2}^{+0.2}$   & ${\phn}1.9\phd_{-0.3}^{+0.4}$   & 31.00 & 31.57 &  0.1 &   3.5 &  16.2 \\ 
 50 &   104522.29-595047.0 &  379 & $22.2\phd_{-0.10}^{+0.06}$ & ${\phn}0.33\phd_{-0.06}^{+0.14}$ & $15.\phd$                       & 32.49 & 30.58 & 11.0 &  -6.1 &  29.9 \\ 
 52 &   104530.22-594821.0 &  218 & $22.2\phd_{-0.2}^{+0.2}$   & ${\phn}2.3\phd_{-0.8}^{+1.5}$    & \nodata                         & 31.50 &   \nodata & 10.3 &  -0.2 &   0.0 \\ 
 55 &   104536.45-594410.7 &  222 & $21.9\phd_{-0.1}^{+0.2}$   & $15.\phd_{-8.5}^{\cdots}$        & \nodata                         & 31.57 &   \nodata &  4.9 &   2.2 &   2.6 \\ 
 56 &   104536.75-594702.2 &  220 & $22.4\phd_{-0.06}^{+0.07}$ & ${\phn}0.58\phd_{-0.1}^{+0.1}$   & \nodata                         & 32.38 &   \nodata & 14.1 &  -5.5 &  13.9 \\ 
 57 &   104538.70-600426.5 &  150 & $21.5\phd_{-0.5}^{+0.3}$   & ${\phn}2.4\phd_{-0.7}^{+1.3}$    & \nodata                         & 31.07 &   \nodata &  2.1 &   0.2 &   0.1 \\ 
 67 &   104615.19-593217.6 &  133 & $20.9\phd_{\cdots}^{+1.0}$ & ${\phn}0.59\phd_{-0.4}^{+0.2}$   & ${\phn}2.1\phd_{-0.5}^{+1.0}$   & 30.64 & 30.92 &  0.5 &   1.8 &   1.3 \\ 
 75 &   104735.26-602923.4 &  199 & $21.9\phd_{-0.3}^{+0.1}$   & ${\phn}0.43\phd_{-0.08}^{+0.16}$ & ${\phn}1.5\phd_{-0.3}^{+1.0}$   & 31.72 & 31.01 &  4.7 &  -2.1 &   2.8 \\ 
 88 &   104858.62-595057.4 &  288 & $20.9\phd_{\cdots}^{+0.6}$ & ${\phn}0.45\phd_{-0.10}^{+0.10}$ & ${\phn}1.3\phd_{\cdots}^{+9.6}$ & 31.15 & 30.45 &  0.5 &   2.0 &   5.3 \\ 
\hline
94 &    104220.83-590908.6 & 511 & $21.3\phd_{-0.4}^{+0.2}$ & ${\phn}0.58\phd_{-0.11}^{+0.09}$ & \nodata & 31.85 & \nodata & 1.2 & 1.2 & 3.9   
\enddata
\tablecomments{Explanations for columns (2), (7), and (8) are given in the footnotes to Table~4.}
\end{deluxetable*}

%
\begin{figure}
\epsscale{1.0}
\plotone{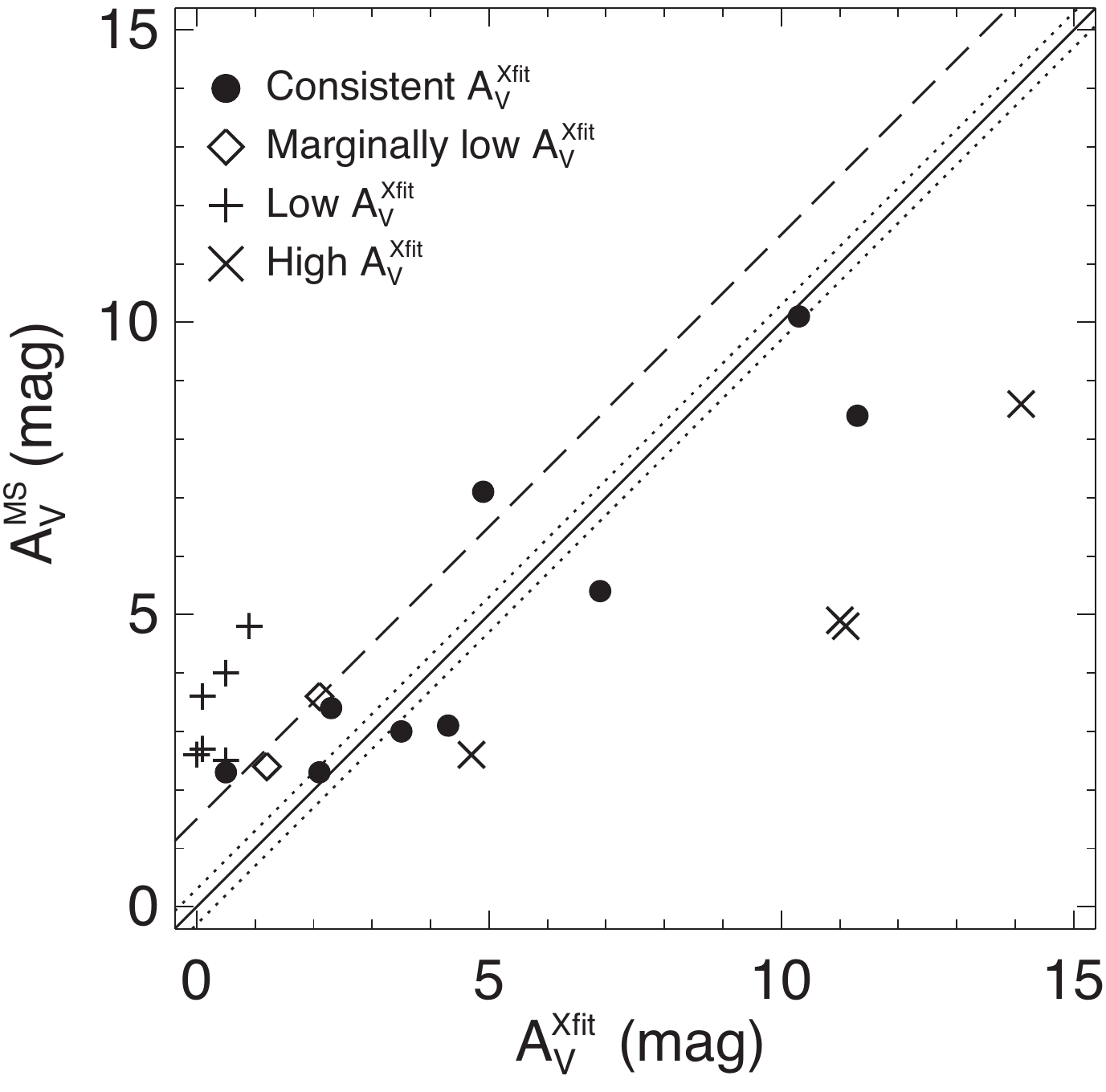}
\caption{Plot of maximum interstellar extinction $A_V^{\rm MS}$ from the IR SED fitting
  versus extinction $A_V^{\rm Xfit}$ from the best-fit absorption $N_{\rm H}^{\rm Xfit}$ for
  the 21 candidate OB stars with  high S/N X-ray spectra
  (Table~5). Diagonal lines represent the divisions between
  the different cases for comparing the IR and X-ray absorption
  results,(see text): $A_V^{\rm Xfit}=A_V^{\rm MS}$ (solid),
  $|A_V^{\rm Xfit}-A_V^{\rm MS}|\le 0.3$~mag (dotted), and $1.5~{\rm
    mag}\ge A_V^{\rm Xfit}-A_V^{\rm MS}>0.3$~mag (dashed).
\label{avxfit}}
\end{figure}

Table~5 includes comparisons between extinction
$A_V^{\rm Xfit}$, derived from the best-fit X-ray
absorption $N_{\rm H}^{\rm Xfit}$, 
and $A_V^{\rm MS}$
(columns 9 and 10), and these comparisons are plotted in
Figure~\ref{avxfit}. Column (11) gives the change $\Delta C_{\rm stat}$ in the 
goodness-of-fit parameter between the frozen-absorption fits of
Table~4 and the best free-absorption fit. XSPEC defines a 90\% confidence
interval as having $\Delta 
C_{\rm stat}<2.72$, hence for the 9 sources in Table~5 meeting this
criterion, we consider the X-ray spectral fitting results to be
consistent with the IR SED fitting results. These sources, plotted as
filled circles in Figure~\ref{avxfit}, are OBc 2, 12, 19,
  30, 39, 52, 55, 57, and 67.

Based on the well-behaved degeneracy between $T_{\rm  
    eff}$ and $A_V$ (Figure \ref{HRD}) and the ${\sim}0.3$~mag uncertainty on $A_V$
when only the IR SED is used in the fitting (Figure \ref{IRonly}) we
can divide the 21 high-S/N sources into the following cases (see Figure~\ref{avxfit}):
\begin{itemize}
\item $|A_V^{\rm MS}-A_V^{\rm Xfit}|\le 0.3$~mag. The X-ray spectral
  fitting results are consistent with a single OB star on or near the MS.
    OBc 2, 52 and 57 are in this group, and all are also consistent with the
  ``frozen'' fits according to $\Delta C_{\rm stat}$ (Table~5).
\item $1.5~{\rm mag}\ge A_V^{\rm MS}-A_V^{\rm Xfit} > 0.3$~mag. X-ray
  spectral fitting suggests an absorption lower than the MS
  intersection point but still consistent with $10^4~{\rm
    K}<T_{\rm eff}<T_{\rm eff}^{\rm MS}$. OBc 12 and 94 are in this
  group, meaning these stars are likely massive Carina members but may
  be unresolved, 
  binary systems or post-MS (super)giants. 
\item $A_V^{\rm MS}-A_V^{\rm Xfit} \ge 1.5$~mag. The X-ray spectrum
  includes a soft, unabsorbed component, and hence the spectral
  fitting results are inconsistent with the high absorption required
  by the SED fitting results.
This could cast doubt on the OB candidacy of these stars,
since lower absorption correlates with cooler stars (Figure~\ref{HRD}b).
 OBc 1, 10, 18, 32, 34, 41,
and 88 fall into this category (as would OBc 55 and 67, except
they are still within the 90\% $\Delta C_{\rm stat}$ confidence interval; Table~5).
\item $A_V^{\rm MS}-A_V^{\rm Xfit} < -0.3$~mag. X-ray spectral
  fitting prefers significantly higher absorption than the maximum
  allowed by IR SED fitting. There are 2 possible explanations, both
  of them consistent with X-rays originating from an OB star:
  X-rays are absorbed ``locally'' in the stellar wind
  \citep[][]{OBglobal}, or a degeneracy 
  in the XSPEC fitting introduces a spurious, highly-luminous but
  highly-absorbed soft thermal plasma (the latter is more
  likely in cases where deviation is 
  more than a few mag). OBc 5, 50, 56, and 75 fall into this category
  (as would OBc 19, 30, and 39, except they are still
  within the 90\% $\Delta C_{\rm stat}$ confidence interval.) 
\end{itemize} 
In summary, the XSPEC fitting results are consistent with reddened OB
stars in the majority of the 21 high-S/N cases (Table~5 and
Figure~\ref{avxfit}). 
We might wonder 
if the 7 stars for which XSPEC prefers very low absorption
(listed above in the 3rd bullet 
point) are foreground stars instead of Carina OB stars, however this
category includes 5 stars with remarkable X-ray 
emission that seems very unlikely to originate from a low-$L_{\rm bol}$
star that happens to be located in front of the Carina Nebula. OBc 1,
10, 18, and 32 all boast ${>}1000$ net counts, making them among the
brightest sources in the entire CCCP catalog
\citep{CCCPcatalog}. OBc 41 was observed to produce a spectacular
X-ray flare with a peak luminosity $\log{L_{t,c}}\ge
33.9$~erg~s$^{-1}$ \citep[assuming this object 
is at the Carina distance;][]{CCCPintro}; the spectral fitting results
for OBc 41 presented in Tables~4 and 5 correspond to the
quiescent spectrum.


\section{Discussion and Summary}

We have identified 94 candidate X-ray-emitting OB stars in the Carina Nebula.
The majority have not been identified previously, probably because
they tend to be more obscured and located outside of
well-studied, ionizing clusters. The average extinction (upper limit)
among the candidate OB stars is $\overline{A}_V^{\rm MS}=5.8$~mag,
compared to $\overline{A}_V=1.9$~mag among 
the known OB stellar population.

Using a validation sample of 182 OB stars in the Carina Nebula with
known spectral types, we demonstrate that optical+IR SED fitting provides a
robust method for simultaneously measuring the bolometric luminosity
and extinction of individual stars. While not significant in the
current analysis, systematic errors
introduced by the inappropriateness of the ATLAS9 model atmospheres
\citep{Kurucz} when applied to early O stars and OB supergiants
should motivate the incorporation of fully line-blanketed, non-LTE,
expanding atmosphere models into the SED fitting analysis for future
applications. 

We find that the extinction law measured toward the OB stars has two
components: $A_V=1$--1.5~mag produced by foreground dust in the 
diffuse interstellar medium (ISM) plus a contribution from local dust in
the Carina molecular clouds that increases with as $A_V$ increases. In
other words, the ``anomalous'' extinction law is more readily observed
for stars located deeper inside or behind the obscuring Carina molecular
clouds. The local dust 
component is characterized by a ratio of total-to-selective
absorption $R_V>4.0$.
While a subset of the least-reddened stars 
agrees with the standard diffuse ISM extinction law ($R_V=3.1$), the {\it
  average} extinction law toward the 
set of known Carina OB stars is better represented by $R_V=4.0$.
Similar two-component extinction laws, with higher $R_V$ values produced by
local dust, have been measured toward other Galactic
\hii regions, for example M17 \citep{CW98,VH08} and NGC 3603 \citep{AP00,NM08}.


Higher $R_V$ values generally correspond to dust grain distributions with
larger average grain sizes \citep[e.g.][]{CCM89,DW01,FM09}. \citet{WH76}
proposed that smaller dust grains are preferentially evaporated in the
harsh radiation field permeating the Carina \hii region, weighting
the grain distributions toward larger sizes. More recent work, however, suggests
that dust grains are rapidly destroyed and replenished inside
energetic \hii regions, 
but relatively small column densities
are sufficient to produce the bright mid-IR emission often observed to be
coincident with the ionized gas
\citep{EC10}. Dust processed within 
the Carina \hii region
therefore may not contribute significantly to the total
line-of-sight extinction.
A more plausible explanation for the ``anomalous'' extinction law in the Carina
molecular cloud could be coagulation of grains or the growth of icy
grain mantles inside cold molecular cloud
fragments, where dust
remains shielded from the external radiation field \citep[e.g.][]{DW01}.
Such cold clouds appear to obscure many regions of the Carina
Nebula. Hot, diffuse plasma produced by the massive stellar population
fills the Carina \hii region cavities, and there is evidence that this
plasma is interacting directly with the molecular clouds, eroding them
\citep{CCCPdiffuse}. Such erosion 
could provide a mechanism for liberating large grains from the cold, dense
molecular clouds, which would explain the widespread impact of this
grain population on the
local extinction law throughout the Carina complex.

While none of the candidate OB stars shows significant IR excess, some
may nevertheless be
as young as, or even younger than, lower-mass YSOs with circumstellar
disks. In the South Pillars region, OB stars are spatially  
intermingled with intermediate-mass YSOs exhibiting IR excess emission and frequently
associated with compact groups of obscured X-ray sources 
\citep[][P11]{spitzcar,CCCPclusters}.  
The spatial correlation of diskless OB
stars with disk-bearing YSOs (Figure~\ref{image}) lends support to the scenario of rapid
disk/envelope destruction 
among intermediate- to high-mass YSOs proposed by P11. 

The most luminous candidate OB stars have
$\log{L^{\rm MS}_{\rm bol}/L_{\sun}}\ga 5.6$, equivalent to O4 V
stars. 
The Tr 16-SE obscured cluster \citep{S07}  
harbors several of these luminous stars, making it a significant
feature of the Carina young stellar population, possibly more massive
than the well-known Bochum 11 or Treasure Chest clusters. 

If confirmed by spectroscopic follow-up, the 94 candidate OB stars
could increase the number of cataloged massive stars in Carina by
30\% to 50\%, depending on the amount of contamination in the sample from
unassociated stars (\S\ref{contam}).  
Of the 200 known massive stars, 140 have 
$\log{L_{\rm bol}/L_{\sun}} \ge 4$ (G11); the remainder have spectral
types of B1 V or later and are less luminous than the candidate OB
stars in our sample. Of these 140, our blind search detected
84, or 60\%, 
because some stars failed to meet the criteria
for inclusion in the CCCP X-ray source catalog \citep{CCCPcatalog}, and some
are located in dense cluster centers, in Tr 16 near $\eta$~Car, in
the Treasure Chest, or
in similar regions of Carina where the 2MASS and IRAC observations are
incomplete due to crowding and/or bright nebular emission. 
Correcting for these factors of incompleteness, we estimate the
Carina Nebula contains between $140+57/0.6=235$ (maximum contamination
case) and
$140+94/0.6=295$ (negligible contamination case) O and early B stars with
$\log{L_{\rm bol}/L_{\sun}}\ge 4$.  
This finding potentially doubles the size of the young massive stellar population
in one of the most spectacular star-forming regions in the Galaxy.


\acknowledgments
We thank the anonymous referee for a prompt, insightful review that
helped improve this work, in particular the discussion 
of systematic errors. We also thank N. R. Walborn and L. M. Oskinova
for helpful conversations. 
M.S.P. is supported by an NSF Astronomy and Astrophysics Postdoctoral
Fellowship under award AST-0901646. 
This work is also supported by
{\it Chandra X-ray Observatory} grant GO8-9131X (PI: L.K.T.) and by
the ACIS Instrument  
Team contract SV4-74018 (PI: G.\ Garmire), issued by the {\it Chandra}
X-ray Center, which is operated by the Smithsonian Astrophysical
Observatory for and on behalf of NASA under contract NAS8-03060.
This work is based on
observations from the {\it Spitzer Space Telescope} GO program 40791
(Vela--Carina; PI: S.R.M.), supported by NASA through an award issued by
the Jet Propulsion Laboratory, California 
Institute of Technology.
This publication makes use of data products from the Two Micron
All-Sky Survey, which is a joint project of the University of
Massachusetts and the Infrared Processing and Analysis
Center/California Institute of Technology, funded by NASA and the NSF.

Facilities: \facility{Spitzer (IRAC)}, \facility{CTIO:2MASS},
\facility{CXO (ACIS)}

\clearpage
\LongTables
\begin{deluxetable}{l@{\hspace{-0.4cm}}c@{\hspace{-0.35cm}}c@{}c@{\hspace{-0.35cm}}c@{}c@{\hspace{-2mm}}r@{\hspace{-1mm}}r@{\hspace{-1mm}}r@{\hspace{-1mm}}r@{\hspace{-1mm}}r@{\hspace{-1mm}}r@{\hspace{-1mm}}rc}\setcounter{table}{1}
\tablecaption{Validation Sample of Cataloged Carina OB Stars: SED Fitting Results and IR Photometry\label{val}}
\tablewidth{0pt}
\tablehead{
\colhead{(1)\tablenotemark{a}} & \colhead{(2)\tablenotemark{a}} & \colhead{(3)} & \colhead{(4)} & \colhead{(5)} & \colhead{(6)} & \multicolumn{7}{c}{(7)--(13)} & \colhead{(14)} \\
\colhead{Star} & \colhead{$A_V(BV)$} & \colhead{$T_{\rm
    eff}$(ST)\tablenotemark{b}} & \colhead{$\log{L_{\rm bol}}$} & \colhead{$A_V$} & \colhead{} & \colhead{$J$} & \colhead{$H$} & \colhead{$K_S$} & \colhead{[3.6]} & \colhead{[4.5]} & \colhead{[5.8]} & \colhead{[8.0]} & \colhead{} \\
\colhead{Name} & \colhead{(mag)} & {(kK)} & \colhead{(\Lsun)} &  \colhead{(mag)} & \colhead{SSTGLMA\tablenotemark{c}} & \colhead{(mag)} & \colhead{(mag)} & \colhead{(mag)} & \colhead{(mag)} & \colhead{(mag)} & \colhead{(mag)} & \colhead{(mag)} & \colhead{IRE?\tablenotemark{d}}
}\\
\startdata
\scriptsize HD 92607    & \scriptsize 1.14 & \scriptsize 35.0 & \scriptsize 5.21 & \scriptsize 1.34 & \scriptsize         G287.1147-01.0236 & \scriptsize   7.91 & \scriptsize   7.87 &  \scriptsize  7.88 &  \scriptsize  7.92 & \scriptsize   7.97 &   \scriptsize 7.98 &  \scriptsize  8.04 &                 \\
\scriptsize HD 305443   & \scriptsize  1.20 & \scriptsize 21.0 & \scriptsize 3.71 & \scriptsize  1.18 & \scriptsize         G287.2169-01.1326 & \scriptsize  10.23 & \scriptsize  10.22 & \scriptsize  10.25 & \scriptsize  10.22 & \scriptsize   10.11 & \scriptsize   9.66 & \scriptsize   8.78 & \scriptsize            I3,I4 \\
\scriptsize HD 92644    & \scriptsize  0.72 & \scriptsize 24.0 & \scriptsize 4.36 & \scriptsize  0.74 & \scriptsize         G287.1386-00.9831 & \scriptsize   8.84 & \scriptsize   8.90 & \scriptsize   8.95 & \scriptsize   8.95 & \scriptsize    9.00 & \scriptsize   8.93 & \scriptsize   8.99 & \scriptsize                \\
\scriptsize LS 1745     & \scriptsize  0.44 & \scriptsize 21.0 & \scriptsize 3.76 & \scriptsize  0.56 & \scriptsize         G287.3038-01.2513 & \scriptsize   9.98 & \scriptsize  10.01 & \scriptsize  10.03 & \scriptsize  10.12 & \scriptsize   10.06 & \scriptsize  10.02 & \scriptsize   9.81 & \scriptsize                \\
\scriptsize HD 92741    & \scriptsize  0.65 & \scriptsize 24.0 & \scriptsize 4.98 & \scriptsize  0.72 & \scriptsize         G287.3067-01.1132 & \scriptsize   7.29 & \scriptsize   7.35 & \scriptsize   7.35 & \scriptsize   7.36 & \scriptsize    7.47 & \scriptsize   7.42 & \scriptsize   7.41 & \scriptsize                \\
\scriptsize LS 1760     & \scriptsize  2.32 & \scriptsize 21.0 & \scriptsize 4.11 & \scriptsize  2.22 & \scriptsize         G287.3069-01.1006 & \scriptsize   9.56 & \scriptsize   9.40 & \scriptsize   9.32 & \scriptsize   9.32 & \scriptsize    9.24 & \scriptsize   9.25 & \scriptsize   9.27 & \scriptsize                \\
\enddata
\tablenotetext{a}{Columns (1) and (2) are from
  G11.}
\tablenotetext{b}{From the \citet{MSH05} calibrations of $T_{\rm eff}$ versus spectral type for O stars, extended to early B stars by \citet{PC05}. Spectral types were compiled from various sources by G11.}
\tablenotetext{c}{\spitzer\ Vela--Carina Point Source Archive identification.}
\tablenotetext{d}{Source exhibits marginal IR excess emission in the specified IRAC band(s): I2 = 4.5~\um, I3 = 5.8~\um, and I4 = 8.0~\um.}
\end{deluxetable}

\clearpage

\begin{deluxetable}{@{\hspace{-9mm}}c@{\hspace{-9mm}}c@{\hspace{-2mm}}c@{\hspace{-6mm}}c@{\hspace{-6mm}}c@{\hspace{-3mm}}c@{\hspace{-4mm}}c@{\hspace{-8mm}}c@{\hspace{-8mm}}c@{\hspace{-8mm}}c@{\hspace{-8mm}}c@{\hspace{-8mm}}c@{\hspace{-8mm}}c@{\hspace{-5mm}}c}\setcounter{table}{3}
\tablecaption{Candidate X-ray Emitting OB Stars\label{can}}
\tablewidth{0pt}
\tablehead{
\colhead{\tiny (1)} & \colhead{\tiny (2)} &
\colhead{\tiny (3)\tablenotemark{b}} & \colhead{\tiny
  (4)\tablenotemark{b}} & \colhead{\tiny (5)\tablenotemark{b}} &
\colhead{\tiny (6)}  & \multicolumn{7}{c}{\tiny (7)--(13)} & \colhead{\tiny (14)} \\
\colhead{\tiny OBc} & \colhead{\tiny } & \colhead{\tiny $T_{\rm eff}^{\rm MS}$} & \colhead{\tiny $\log{L_{\rm bol}^{\rm MS}}$} & \colhead{\tiny $A_V^{\rm MS}$} & \colhead{\tiny } & \colhead{\tiny $J$} & \colhead{\tiny $H$} & \colhead{\tiny $K_S$} & \colhead{\tiny [3.6]} & \colhead{\tiny [4.5]} & \colhead{\tiny [5.8]} & \colhead{\tiny [8.0]} & \colhead{\tiny Other ID/Sp.T.\tablenotemark{d}} \\
\colhead{\tiny No.} & \colhead{\tiny CXOGNC J\tablenotemark{a}} & \colhead{\tiny (kK)} & \colhead{\tiny (${\rm L_{\odot}}$)} &  \colhead{\tiny (mag)} & \colhead{\tiny SSTGLMC\tablenotemark{c}} & \colhead{\tiny (mag)} & \colhead{\tiny (mag)} & \colhead{\tiny (mag)} & \colhead{\tiny (mag)} & \colhead{\tiny (mag)} & \colhead{\tiny (mag)} & \colhead{\tiny (mag)} & \colhead{\tiny or IRE?\tablenotemark{e}}
}
\startdata   
\tiny\phn1 & \tiny  103909.94-594714.5   & \tiny  40.3 & \tiny  5.4 & \tiny 3.6 & \tiny G286.9928-01.0735 & \tiny   8.61 & \tiny   8.03 & \tiny  7.89 & \tiny   7.95 & \tiny   7.97 & \tiny  7.93 & \tiny   7.90  & \tiny                  \\
\tiny   \phn2 & \tiny 104014.67-595654.4   & \tiny 32.8 & \tiny 4.8 & \tiny 3.0 & \tiny        G287.1891-01.1488 & \tiny   9.51 & \tiny   9.14 & \tiny   9.02 & \tiny   8.98 & \tiny   9.04 & \tiny   9.02 & \tiny   9.06  & \tiny                 \\
\tiny   \phn3 & \tiny 104023.46-595039.0   & \tiny 26.1 & \tiny 4.1 & \tiny 3.2 & \tiny        G287.1549-01.0488 & \tiny  10.59 & \tiny  10.31 & \tiny  10.13 & \tiny  10.04 & \tiny  10.03 & \tiny  10.07 & \tiny  10.08  & \tiny                 \\
\tiny   \phn4 & \tiny 104044.15-594934.8   & \tiny 34.3 & \tiny 4.9 & \tiny 9.8 & \tiny        G287.1844-01.0122 & \tiny  11.38 & \tiny  10.33 & \tiny   9.77 & \tiny   9.30 & \tiny   9.15 & \tiny   9.00 & \tiny   8.87  & \tiny                 \\
\tiny   \phn5 & \tiny 104059.29-592724.9   & \tiny 36.2 & \tiny 5.1 & \tiny 4.9 & \tiny        G287.0350-00.6729 & \tiny   9.50 & \tiny   9.00 & \tiny   8.71 & \tiny   8.64 & \tiny   8.50 & \tiny   8.54 & \tiny   8.52  & \tiny                 \\
\tiny   \phn6 & \tiny 104109.05-594359.3   & \tiny 28.9 & \tiny 4.4 & \tiny 3.4 & \tiny        G287.1856-00.9056 & \tiny  10.28 & \tiny   9.80 & \tiny   9.68 & \tiny   9.70 & \tiny   9.74 & \tiny   9.60 & \tiny   9.62  & \tiny                 \\
\tiny   \phn7 & \tiny 104118.94-593544.5   & \tiny 26.0 & \tiny 4.1 & \tiny 3.7 & \tiny        G287.1379-00.7747 & \tiny  10.87 & \tiny  10.34 & \tiny  10.18 & \tiny  10.03 & \tiny  10.18 & \tiny  10.03 & \tiny  10.25  & \tiny                 \\
\tiny   \phn8 & \tiny 104125.96-592319.6   & \tiny 36.0 & \tiny 5.1 & \tiny 7.2 & \tiny        G287.0518-00.5859 & \tiny  10.37 & \tiny   9.39 & \tiny   9.06 & \tiny   8.77 & \tiny   8.84 & \tiny   8.73 & \tiny   8.58  & \tiny                 \\
\enddata
\tablenotetext{a}{CCCP Catalog identifier \citep{CCCPcatalog}.}
\tablenotetext{b}{Values in columns (3)--(5) were derived from SED fitting under the assumption of single massive stars on the theoretical zero-age main sequence (see text). Formally, the $T_{\rm eff}^{\rm MS}$, $L_{\rm bol}^{\rm MS}$ and $A_V^{\rm MS}$ reported here are upper limits and should be used only for comparison purposes among this sample.}
\tablenotetext{c}{\spitzer\ Vela--Carina Point Source Catalog identifier.}
\tablenotetext{d}{Previous star identifications come from the
  following sources: HD = \citet{HDEC}; Y = \citet{C93}; Cr 228
  (Collinder 228) =  \citet{FMF76} or \citet{TBA80}; IRS =
  \citet{RS87}; MJ = \citet{MJ93}; XMM = \citet{XMM}; Bo 11 (Bochum
  11) = \citet{FM87}; and LS = \citet{LS}. Candidate massive stars or
  candidate low-mass PMS stars \citep{NE03,S07} are indicated by
  spectral types OB or PMS, respectively. NOP means that the star was
  not found in any of optical photometric catalogs listed by VizieR
  \citep{VizieR}. PFD indicates that a candidate OB star is also consistent with a foreground MS classification.}
\tablenotetext{e}{Source exhibits marginal IR excess emission in the specified IRAC band(s): I2 = 4.5~\um, I3 = 5.8~\um, and I4 = 8.0~\um.}
\tablenotetext{f}{OB candidate 41 appears to be responsible for the spectacular X-ray flare discovered by \citet{CCCPintro}.}

\end{deluxetable}

\end{document}